%% file: AGreview.tex
\documentclass[twoside,psfig,lscape,epsf]{article}
\usepackage[latin1]{inputenc}
\usepackage{t1enc}
\usepackage{a4}
\usepackage{tabularx}
\usepackage{epsf}
\usepackage{psfig}

\def\bref{\vspace{4pt}\noindent\hangindent=10mm}

\input{abb.tex}

\def\kkl{KK\&L2000}

\begin{document}

\setcounter{figure}{0}
\setcounter{section}{0}
\setcounter{equation}{0}

\begin{center}
{\Large\bf Cosmological Structures behind the Milky Way}
\\[0.2cm]
Ren\'ee C. Kraan-Korteweg\\[0.17cm] 
Departamento de Astronom\'{i}a, Universidad de Guanajuato,\\
Apartado Postal 144, Guanajuato GTO 36000, M\'exico\\
Astronomy Department, University of Cape Town\footnote{since January 2005},\\
Rondebosch 7700, South Africa\\
kraan@circinus.ast.uct.ac.za, http://mensa.ast.uct.ac.za/$\sim$kraan
\end{center}

\vspace{0.5cm}

\begin{abstract}
\noindent{\it This paper provides an update to the review on
extragalactic large-scale structures uncovered in the Zone of
Avoidance (ZOA) by Kraan-Korteweg \& Lahav 2000, in particular in the
Great Attractor (GA) region. Emphasis is given to the penetration of
the ZOA with the in 2003 released near--infrared (NIR) 2MASS Extended
Source Catalog. A comparison with deep optical searches confirms that
the distribution is little affected by the foreground dust. Galaxies
can be identified to extinction levels of over $A_B \ga 10^{\rm m}$
compared to about $3\fm0$ in the optical. However, star density has
been found to be a strong delimiting factor. In the wider Galactic
Bulge region ($\ell = 0\deg \pm 90\deg$) this does not hold and
optical surveys actually probe deeper (see Fig.~\ref{2M_ZOA}). The
shape of the NIR-ZOA is quite asymmetric due to Galactic features such
as spiral arms and the Bulge, something that should not be ignored
when using NIR samples for studies such as dipole determinations.

Various systematic surveys have been undertaken with radio telescopes
to detect gas-rich galaxies in the optically and NIR impenetratable
part of the ZOA. We present results from the recently finished deep
blind HI ZOA survey performed with the Multibeam Receiver at the 64\,m
Parkes telescope ($v \la 12\,700$~\kms). The distribution of the roughly
one thousand discovered spiral galaxies within $|b| <5\deg$ clearly
depict the prominence of the Norma Supercluster. In combination with
the optically identified galaxies in the ZOA, a picture emerges that
bears a striking resemblance to the Coma cluster in the Great Wall in
the first redshift slice of the CFA2 survey (de Lapparent, Geller \&
Huchra 1986): the rich Norma cluster (ACO\,3627) lies within a
great-wall like structure that can be traced at the redshift range of
the cluster over $\sim 90\deg$ on the sky, with two foreground
filaments -- reminiscent of the legs in the famous stick man -- that
merge in an overdensity at slightly lower redshifts around the radio
galaxy PKS\,1343$-$601 (see Figs.~\ref{wedge5} \& \ref{wedge10}).}
\end{abstract}

\section{Introduction}
\begin{figure}[t!]
\hfil \psfig{figure=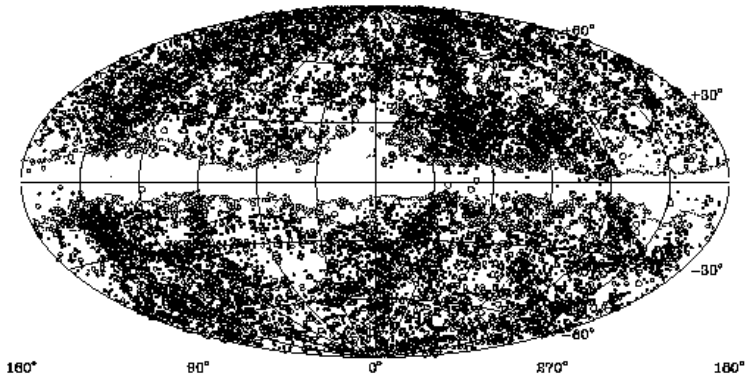,width=10cm} \hfil
\caption
{Aitoff equal-area projection in Galactic coordinates of galaxies with
${D}\ge1\farcm$3. The galaxies are diameter-coded. The contour marks
absorption in the blue of ${A_B}= 1\fm0$ as determined from the
Schlegel, Finkbeiner \& Davis (1998) dust extinction maps. Figure
from \kkl.}\label{ait}\protect
\end{figure}

The absorption of light due to dust particles and the increase in star
density close to the Galactic Equator and around the Galactic Bulge
creates a ``Zone of Avoidance'' in the distribution of galaxies, the
size and shape of which depends on the wavelength at which galaxies
are sampled. Figure~\ref{ait} shows a complete sample of optically
cataloged galaxies in an Aitoff projection in Galactic coordinates
(see Kraan-Korteweg \& Lahav 2000 for details; henceforth \kkl). The
broad band void of galaxies takes up about 20\% of the sky. Its form
is like a near-perfect negative of the optical light distribution as
depicted in the famous composite by Lundmark (1940), and is well
traced by the dust (see contour in Fig.~\ref{ait}).

\begin{figure}[t!!]
\hfil 
\psfig{figure=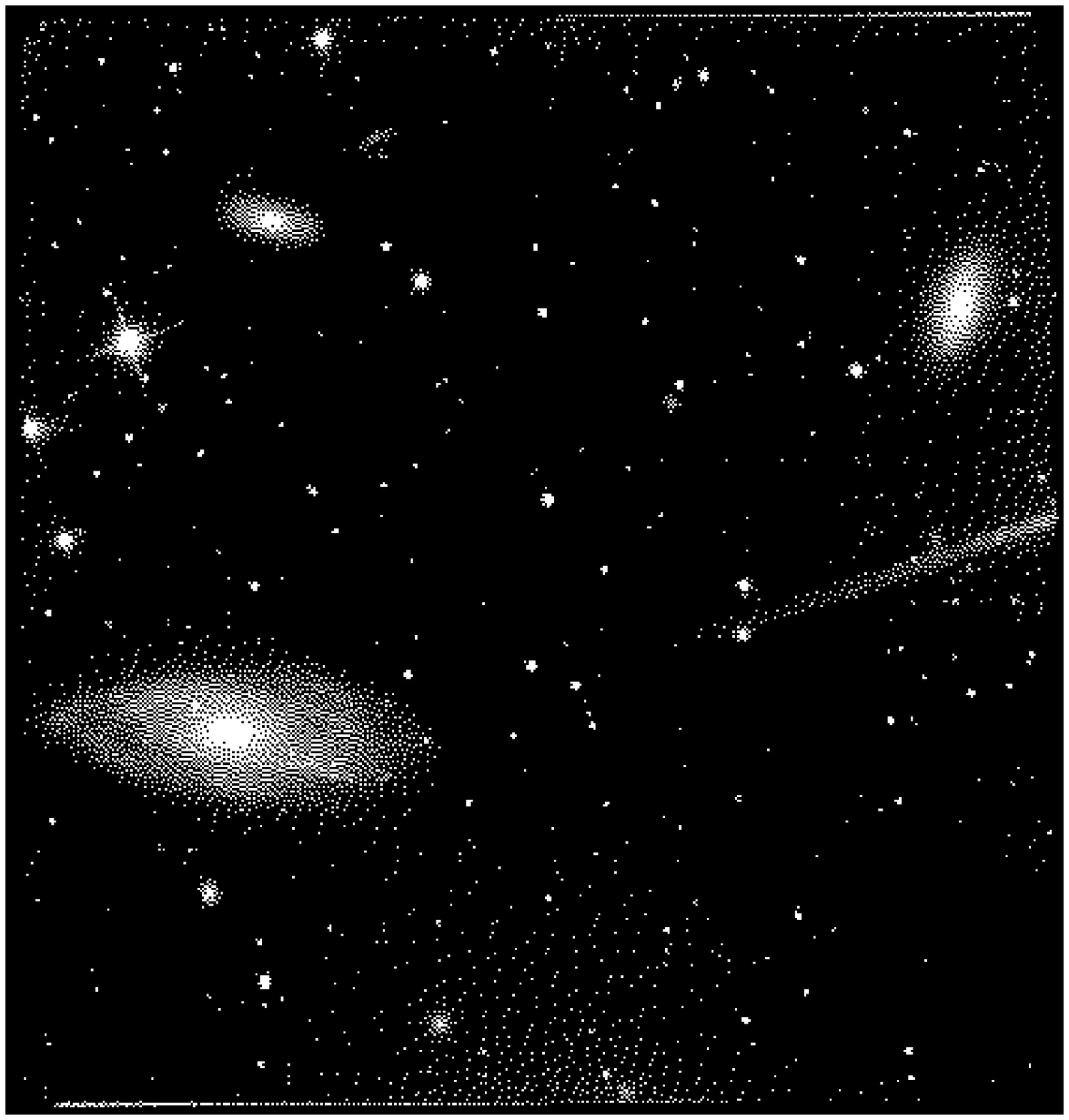,height=3.8cm,width=3.8cm,bbllx=67pt,bblly=167pt,bburx=546pt,bbury=624pt,clip=,angle=-90}
\psfig{figure=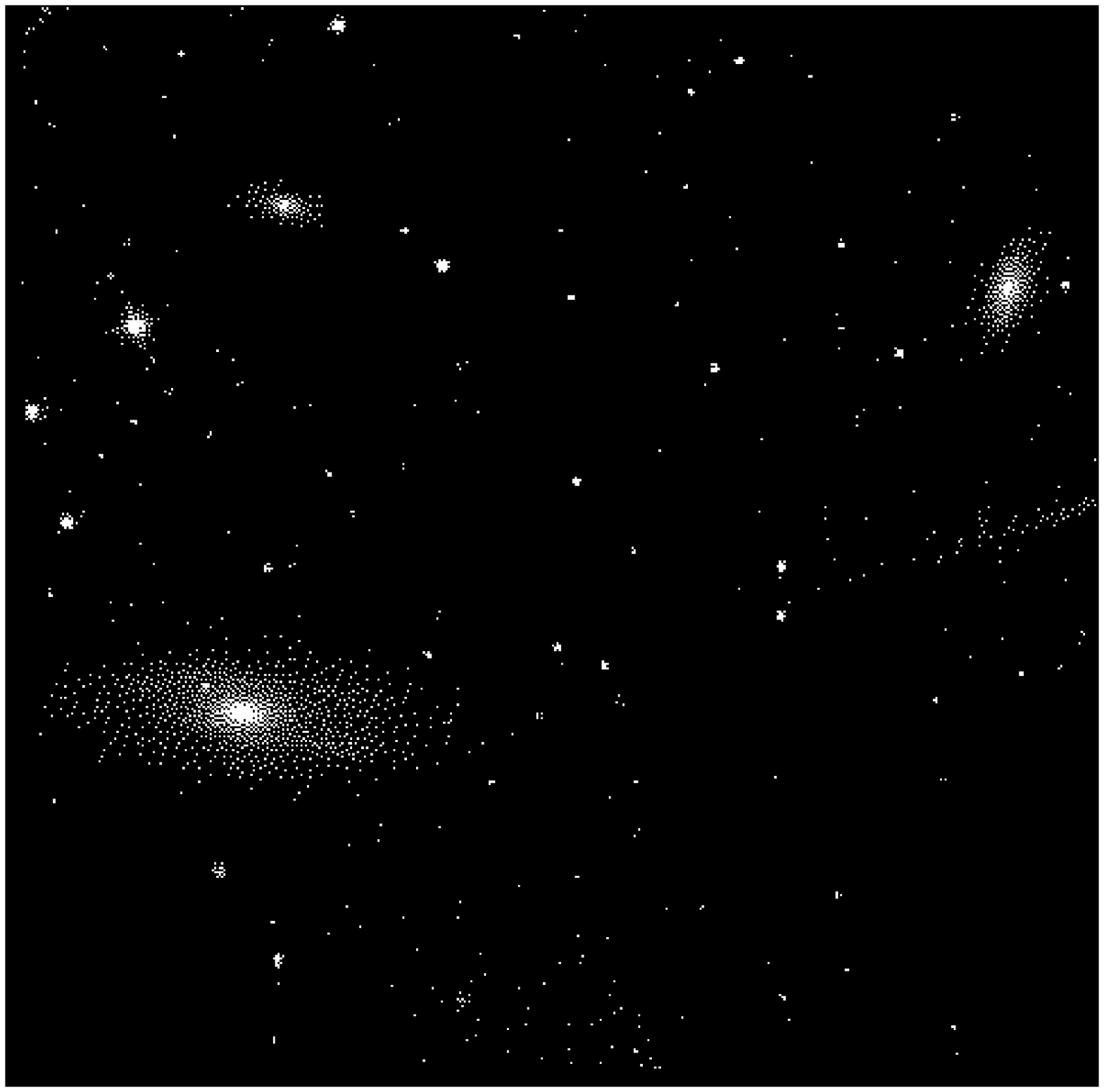,height=3.8cm,width=3.8cm,bbllx=67pt,bblly=155pt,bburx=544pt,bbury=637pt,clip=,angle=-90}
\psfig{figure=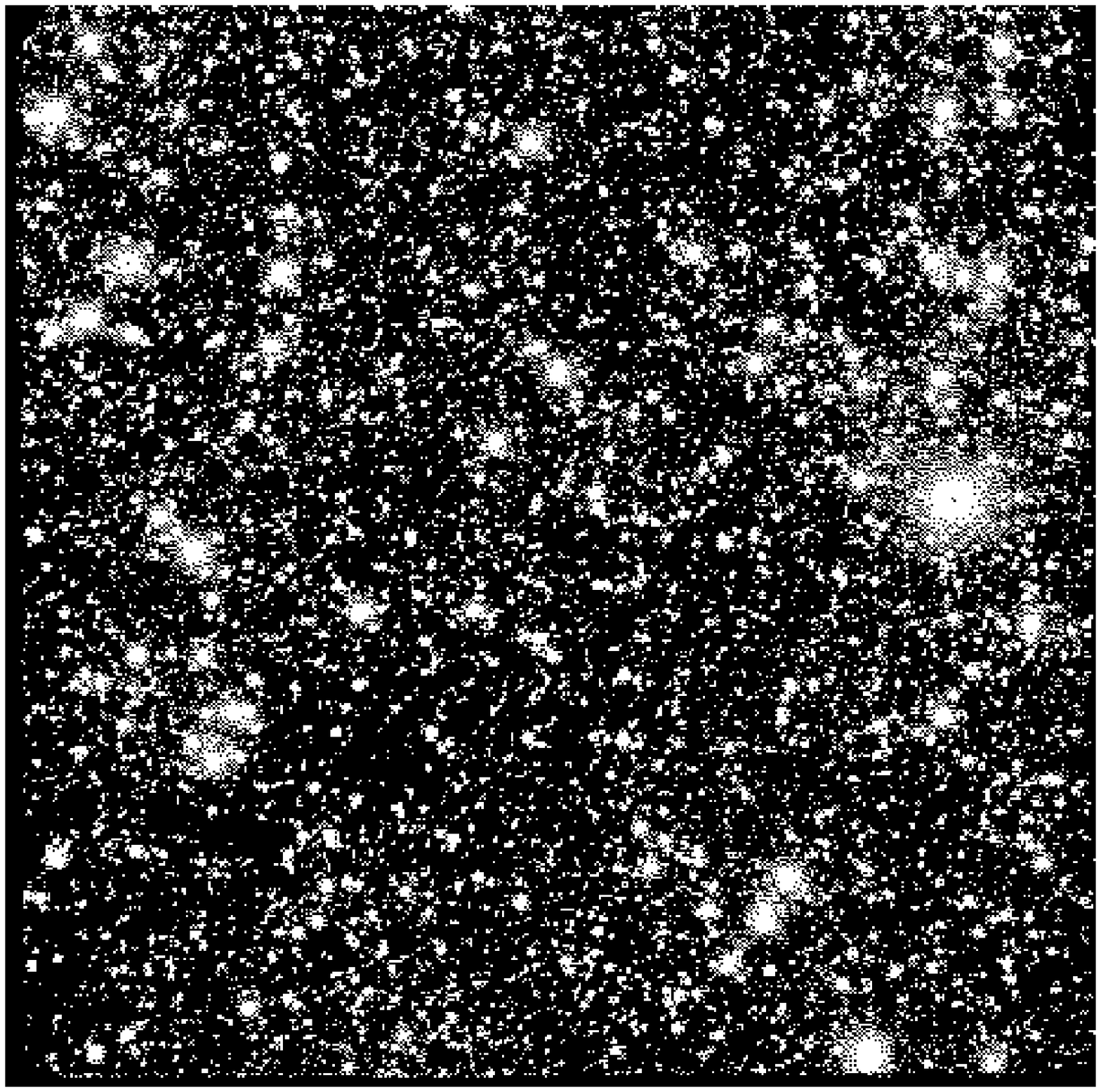,height=3.8cm,width=3.8cm,bbllx=67pt,bblly=154pt,bburx=546pt,bbury=637pt,clip=,angle=-90}
\hfil
\caption
{Left: A $JHK$ image of 8'$\times$8' obtained with the Japanese IRSF
 of the Hydra galaxy cluster. Middle: simulation of same field seen
 through an obscuration layer of 12\fm0, 2\fm5, 1\fm7, and 1\fm1 of
 extinction in the $BJHK$ bands respectively. Right: previous field
 now positioned in the area of the low latitude radio galaxy
 PKS\,1343$-$601. Figure adopted from Nagayama 2004.}
\label{simu}
\end{figure}

The ZOA has, with few exceptions, been avoided by astronomers studying
the extragalactic sky because of the inherent difficulties in
determining the physical parameters of galaxies lying behind the disk
of our Galaxy -- if they can be identified at all.  The effect of
absorption and star-crowding is illustrated in a simulation made by
Nagayama (2004) which is reproduced in Fig.~\ref{simu}. The images are
based on observations made with the Japanese 1.4\,m Infrared Survey
Facility (IRSF) at the Sutherland observing site of the South African
Astronomical Observatory.  The camera on the IRSF has the ability to
simultaneously take $J$, $H$, and $K$ data with a field of view of
$8' \times 8'$.

The left-hand panel shows a combined $JHK$ field in the Hydra cluster
where absorption is negligible and star-crowding a minor
problem. Subjecting this image to a foreground extinction of
$A_B=12^{\rm m}$ which in the NIR bands $JHK$ reduces to a mere
$2\fm5, 1\fm7$ and $1\fm1$ respectively (see Sect.~\ref{NIR} for
details) results in the image of the middle panel. While the
originally small and faint galaxies are lost completely, the larger
galaxies are smaller in size, of lower surface-brightness, and
redder. A further difficulty in identifying galaxies and determining
their properties is star-crowding. To illustrate this, Nagayama then
combined this artificially absorbed field with a field in the
surroundings of the radio galaxy PKS\,1343$-$601 at low Galactic
latitudes ($\ell=309\fdg7, b=+1\fdg8$), also obtained with IRSF. This
is reproduced in the right-hand panel of Fig.~\ref{simu}. Here, the
recognition of even the intrinsically largest galaxies is hard when
knowing where the galaxies are located (see panels to the left).
Further examples of extinction and star-crowding effects can be found
on http://www.z.phys.nagoya-u.ac.jp/$\sim$nagayama/hydra. This site
also shows some examples of real galaxy candidates detected behind
this thick obscuration and star layer.

The simulation clearly demonstrates why the incentive was low to map
and investigate galaxies, their properties and their distribution in
space in the ZOA. However, with the realization that galaxies are
located predominantly in clusters, sheets and filaments, leaving large
areas devoid of luminous matter, came the understanding that a
concensus of galaxies of the ``whole sky'' is required when addressing
various cosmological questions related to the dynamics of the local
Universe.

The closest superclusters such as the Local Supercluster, the
Centaurus Wall, the Perseus-Pisces chain, the Great Attractor --
a large mass overdensity of about $5 \cdot 10^{16}{\cal M}_\odot$
that was predicted from the systematic infall pattern of 400
ellipiticals (Dressler et al.~1987) -- all are bisected by the Milky
Way, making a complete mapping and determination of their extent and
dynamics impossible. Moreover, the irregular distribution of mass
induces systematic flow patterns over and above the uniform Hubble
expansion of the Universe. This effect is seen in the peculiar motion
of the Local Group with respect to the Cosmic Microwave Background
(CMB; e.g. Kogut et al. 1993). Such systematic flow patterns were first
mapped within the Virgo Supercluster (Tonry \& Davis 1981) and later
on a much larger scale later in the Great Attractor region. It might
even perturb the motions of galaxies in a volume all the way out to
the Shapley Concentration, including the GA as a whole, though this
still remains controversial (e.g. Kocevski, Mullis, \& Ebeling 2004;
Lucey, Radburn-Smith \& Hudson, 2005; Hudson et al. 2004).

Kolatt, Dekel \& Lahav (1995), have shown that the mass distribution of
the inner ZOA ($b \pm20\deg$) as derived from theoretical
reconstructions of the density field is crucial to the derivation of
the gravitational acceleration of the LG. This not only concerns
hidden clusters, filaments, and voids. Nearby massive galaxies may
also contribute significantly to the dipole and many of the nearby
luminous galaxies do actually lie behind the Milky Way
(e.g. Kraan-Korteweg et al. 1994). Moreover, a hidden Andromeda-like
galaxy will influence the internal dynamics of the LG, its mass
derivation and the present density determination of the Universe from
timing arguments (Peebles 1994).

For our understanding of velocity flow fields, in particular the Great
Attractor, the ZOA constitutes a severe barrier. The various 2 and
3-dimensional reconstruction methods find the flow towards this GA to
be due to a quite extended region of moderately enhanced galaxy
density centered on the Milky Way (about $\sim 40\deg \times 40\deg$,
centered on $\ell,v,b \sim 320\deg, 0\deg, 4500$\kms; see e.g. Fig.~1b
in Kolatt et al.~1995). Although a considerable excess of galaxies is
seen in that general region of the sky (Lynden-Bell \& Lahav 1988;
Fig.~1 here), no dominant cluster or central peak had been
identified. Whether it existed and whether galaxies were fair tracers
of the dynamically implicated mass distribution could not be answered,
because a dominant fraction of the GA was hidden by the Milky Way.

For these reasons, various groups began projects in the last 10--15
years to try to unveil the galaxy distribution behind our Milky
Way. Preliminary results, mainly based on optical, HI observations and
follow-up of selected IRAS-PSC galaxy candidates, were presented at
the first conference on this topic in 1994 (see Balkowski \&
Kraan-Korteweg (eds.) 1994, ASP Conf. Ser. 67).  Meanwhile most of the ZOA
has been probed in a ``systematic'' manner in ``all'' wavelength
ranges of the electromagnetic spectrum (optical, near- and
far-infrared, HI and X-ray), next to, and in comparison to, the
reconstructed density fields in the ZOA. Many of these surveys and
subsequent results are described in the proceedings of the second and
third meeting on this topic (Kraan-Korteweg, Henning \& Andernach
(eds.) 2000, ASP Conf. Ser. 218; Fairall \& Woudt (eds.) 2005, ASP
Conf. Ser. 329).

A comprehensive overview on the then current status of all the ZOA
projects was prepared in 2000 by Kraan-Korteweg \& Lahav.  It provides
a detailed introduction on the motivation of ZOA studies, the status
of and the results from the different survey methods, including a
discussion on the limitations and selection effects of the various
approaches, as well as how they complement each other.  In this paper,
I will build on the information given there, and concentrate mainly on
new results, although a summary and an update on the results from deep
optical galaxy searches and redshift follow-ups in the Great Attractor
region is given in Sect.~\ref{GA}, as these are relevant to the
discussions in the subsequent sections. Sect.~\ref{NIR} describes the
enormous progress made in NIR-surveys with the release of the 2MASS
Extended Source Catalog (2MASX) which contains 1.65 million galaxies
or other extended sources over the whole-sky. It contains a discussion
on the characteristics of this survey, in particular to penetrating
the ZOA in comparison to optical surveys.  Sect.~\ref{HI} is dedicated
to the results obtained from the systematic HI-survey of the
southern ZOA that was performed between 1997 and 2002 with the
Multibeam Receiver at the 64\,m Parkes radio telescope. Next to the
instrument and survey technique, the newly uncovered galaxy
distribution is discussed. The last section (Sect.~\ref{Sum}) will
then describe the emerging picture of the Great Attractor overdensity
including the data obtained from the various ZOA survey methods.

\section{Optical Galaxy Searches and the GA}\label{GA}

Optical galaxy catalogs become increasingly incomplete for
intrinsically large galaxies towards the Galactic Plane, because of
the reduction in brightness and `visible' extent by the thickening
dust layer and the increase in star density
(Fig.~\ref{simu}). However, this is a gradual effect. Deeper searches
for partially obscured galaxies -- fainter and smaller than existing
catalogs -- have been succesfully performed on existing sky survey
plates. Over 50\,000 unknown galaxies were uncovered, resulting in a
considerable reduction of the ZOA.

\begin{figure}[ht]
\hfil \psfig{figure=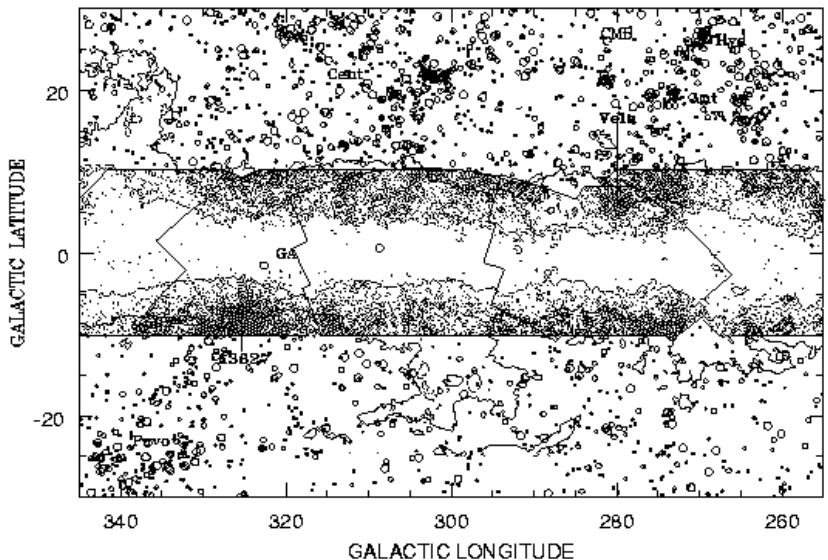,width=11cm,bbllx=28pt,bblly=161pt,bburx=566pt,bbury=
532pt,clip=,angle=0} \hfil
\caption
{Distribution of Lauberts (1982) galaxies with $D \ge 1\farcm3$ (open
circles) and galaxies with $D \ge 12\arcsec$ (small dots) identified
in the optical galaxy searches. The contours represent extinction
levels of $A_B = 1\fm0$ and $3\fm0$.}
\label{WKK}
\end{figure}

Fig.~\ref{WKK} gives an example of results obtained by our group in
and around the Great Attractor region. This comprises (from right to
left) the Vela region (Salem \& Kraan-Korteweg, in prep.),
Hydra/Antlia (Kraan-Korteweg 2000), Crux and Great Attractor (Woudt \&
Kraan-Korteweg 2001) and Scorpius (Fairall \& Kraan-Korteweg 2000,
2005).  Using a viewer with a 50 times magnification on IIIaJ film
copies of the ESO/SRC survey resulted in the identification of over
17\,000 galaxies down to a diameter limit of $D = 0\farcm2$ within
Galactic latitudes of $|b| \la 10\deg$ over a longitude range of
$250\deg \la \ell \la 350\deg$. 97\% were previously unknown. Their
distribution is displayed in Fig.~\ref{WKK} together with earlier
known galaxies. Note how the ZOA could be filled from $A_B =
1\fm0$, the approximate completeness limit of previous catalogs, to
$A_B = 3\fm0$, with a few galaxies still recognizable up to extinction
levels of $A_B = 5\fm0$.

Distinct large-scale structures uncorrelated with the foreground
obscuration can be discerned. The most extreme overdensity $(\ell,b)
\sim (325\deg,-7\deg$) is found close to the core of the GA. It is
centered on the cluster ACO\,3627 (Abell, Corwin \& Olowin 1989) and
is at least a factor 10 denser compared to regions at similar extinction
levels, and contains a significant larger fraction of brighter
galaxies as well as elliptical galaxies. The prominance of this
cluster had never been realized because of its location in the Milky
Way. Within the Abell radius (defined as 3 $h_{50}^{-1}$ Mpc) of this
cluster, a total of 603 galaxies with $D \ge 0\farcm2$ have been
identified, of which only 31 were cataloged before by Lauberts
(1982).

\begin{figure}[h]
\hfil \psfig{figure=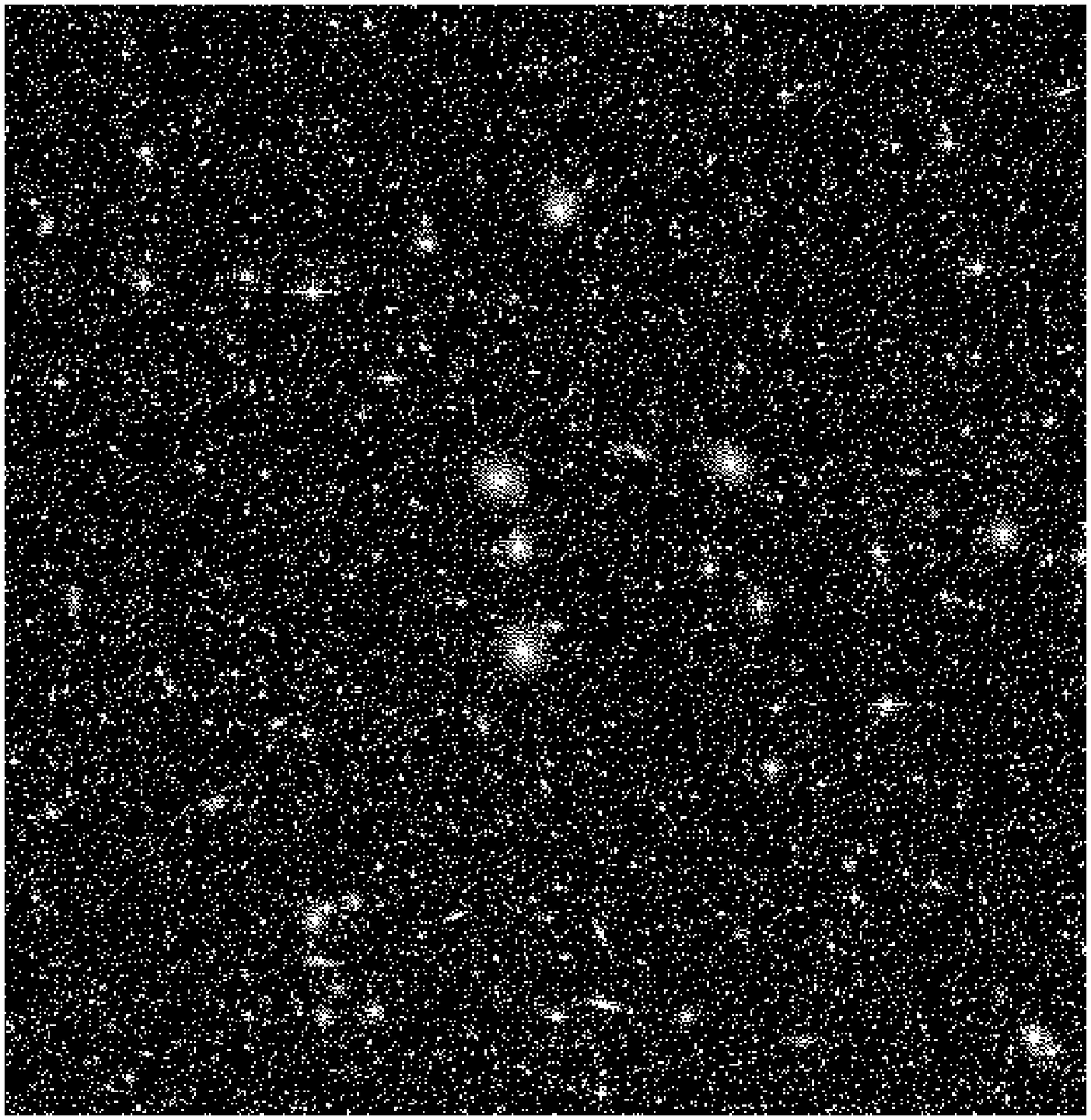,height=5.5cm,width=5.5cm,bbllx=36pt,bblly=133pt,bburx=575pt,bbury=658pt,clip=,angle=-90} \hspace{0.1cm} 
      \psfig{figure=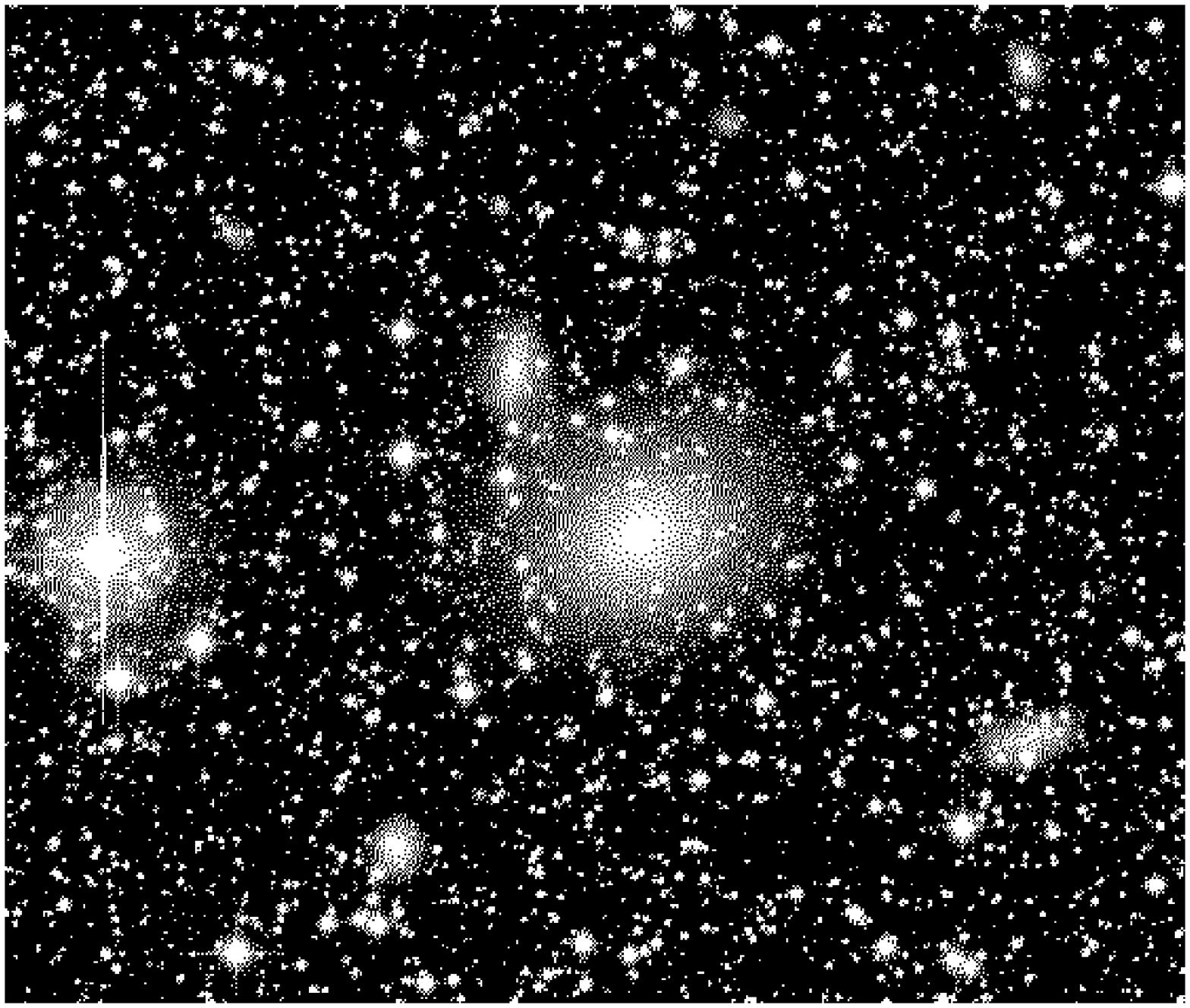,height=5.5cm,width=5.5cm,bbllx=36pt,bblly=161pt,bburx=575pt,bbury=634pt,clip=,angle=-90} \hfil
\caption{A $BRI$ composite of the central $37\arcmin \times 37\arcmin$
of the cluster ACO\,3627 obtained with the WFI of the ESO 2.2\,m
telescope (left). Note the 2 dominant cD galaxies at the center of
this cluster and the richness of the dwarf population in the close-up
(4x) on the right. Figure from Woudt et al. 2000.}
\label{norma}
\end{figure}

Figure~\ref{norma}, a deeper $BRI$ composite image obtained later by
Woudt with the Wide Field Imager at the ESO 2.2\,m telescope in la
Silla (Woudt, Kraan-Korteweg \& Fairall 2000), displays the center of
this cluster in its full glory, despite the approximate 200\,000
foreground stars.  On this $37\arcmin \times 37\arcmin$ image (left
panel), even more galaxies can be identified than the 74 galaxies
found in the same area on the IIIaJ fields. The cluster has, like the
Coma cluster, two cD galaxies at its center. The richness of this
cluster can be fully appreciated in the close-up in the right-hand panel 
with the there apparent large dwarf galaxy population (image centered
on the right of center cD of the left panel).

A quantification of the relevance of the newly uncovered galaxy
distribution in context to known structures can be made when studying
the completeness limits of the ZOA galaxy catalogs and correcting the
observed parameters of the galaxies for the diminishing effects due to
absorption applying the inverse Cameron (1990) laws. The latter
provides equations to correct the observed magnitudes and diameters
for their reduction as a function of extinction at their positon
behind the Milky Way.  Kraan-Korteweg (2000) and Woudt \&
Kraan-Korteweg (2001) have shown that their catalogs are complete to
$D=14\arcsec$ down to extinction levels of $A_B \le
3\fm0$. Corrections for the apparently smallest galaxy at maximum
extinction indicates that these surveys are complete for galaxies with
extinction-corrected diameters of $D^o = 1\farcm0$ down to $A_B \le
3\fm0$.

\begin{figure}[h]
\hfil \psfig{figure=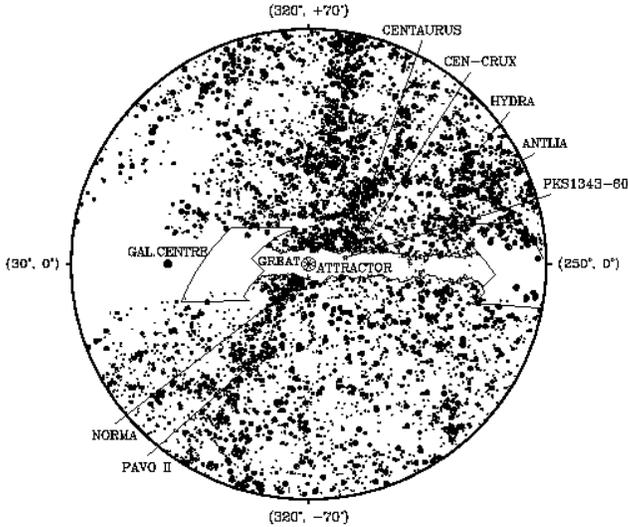,width=8.5cm,bbllx=64pt,bblly=234pt,bburx=578pt,bbury=659pt,clip=,angle=0} \hfil
\caption{Equal area projection of galaxies with $D^0 \ge 1\farcm3$ and
$A_B \le 3^{\rm m}$ centered on the GA at $(\ell, b) = (320\deg,
0\deg)$ within a radius of 70$\deg$. The galaxies are taken from the
ESO, the UGC and the MGC, complemented by our ZOA catalogs. The
extinction contour of $A_B = 3^{\rm m}$ is superimposed.  Search areas
in progress are indicated. Figure from Woudt \& Kraan-Korteweg 2001.}
\label{ga}
\end{figure}

Knowing that the combined ESO Lauberts (1982) catalog, the Uppsala
General Catalog UGC (Nilson 1973), and the Morphological Catalog of
Galaxies MGC (Vorontsov-Velyaminov \& Archipova 1963-74) displayed in
Fig.~\ref{ait} are complete to $D = 1\fdg3$ (Hudson \& Lynden-Bell
1991), we can complement this merged whole-sky catalog down to $A_B
\le 3\fm0$ with galaxies that would appear in these catalogs were they
not lying behind the Milky Way, i.e. with $D^o = 1\farcm3$. This has
been done in Fig.~\ref{ga}, in an equal area projection centered on the
Great Attractor at $(\ell, b) = (320\deg, 0\deg)$ and includes all
galaxies with extinction-corrected diameters larger than $D^0 \ge
1\farcm3$ and $A_B \le 3^{\rm m}$ from the Hydra/Antlia, Crux and GA
catalogs, next to the previously known galaxies from the ESO, UGC and
MGC catalogs.

Figure~\ref{ga} provides the most complete view of the optical galaxy
distribution in the Great Attractor region to date.  Comparing
Fig.~\ref{ga} with Fig.~\ref{ait} demonstrates the reduction of the
optical ZOA (over 50\%).  The final galaxy distribution not only shows
the dominance of the ACO\,3627 cluster, henceforth called the Norma
cluster for the constellation in which it is located, but also
displays a high galaxy density in the wider GA region on both sides of
the Galactic Plane. Though the reduction of the ZOA is significant,
optical approaches clearly do not fully succeed in penetrating the
ZOA.

\subsection{Velocity Distribution}\label{vel}

Redshift coverage of the nearby galaxy population in the ZOA is
essential in determining their impact on the local dynamics and also
to identify the features that form part of the GA overdensity.  We
have aimed to obtain a fairly homogeneous and complete coverage of the
(extinction-corrected) brighter and larger newly uncovered galaxies in
the ZOA. Three distinctly different observational approaches were
used: (i) optical spectroscopy for individual galaxies of high central
surface brightness at the 1.9\,m telescope of the SAAO
(Kraan-Korteweg, Fairall, \& Balkowski 1995; Fairall, Woudt, \&
Kraan-Korteweg 1998; Woudt, Kraan-Korteweg, \& Fairall 1999), (ii)
HI-line observations with the 64\,m Parkes radio telescope for low
surface brightness gas-rich spiral galaxies (Kraan-Korteweg, Henning
\& Schr\"oder 2002; Schr\"oder, Kraan-Korteweg \& Henning, in prep.),
(iii) low resolution, multi-fiber spectroscopy for the high-density
regions with Optopus and MEFOS at the 3.6\,m telescope of ESO, LaSilla
(see Woudt et al. 2004 for MEFOS results).  With the above
observations, we typically obtain redshifts of 15\% of the galaxies
and can trace large-scale structures fairly well out to recession
velocities of about $20\,000$\kms.

\begin{figure}[ht]
\hfil \psfig{figure=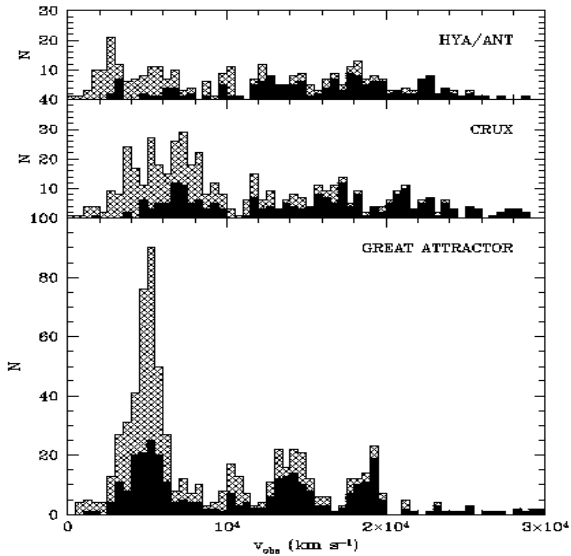,height=7.5cm,width=7.5cm} \hfil
\caption{Velocity histograms in the Hydra/Antlia, Crux and GA
regions. Dark shaded area correspond to the heliocentric velocities
obtained with MEFOS, cross-hatched region includes the SAAO and Parkes
observations as well as velocities from the literature.  Figure from
Woudt et al. 2004.}
\label{mefos}
\end{figure}

The resulting velocity histograms are plotted in Fig.~\ref{mefos},
separately for the three search areas.  One glance immediately reveals
the striking difference between the Hydra/Antlia, Crux and GA region,
although all three have been sampled to approximately the same depth.
The velocity distribution in Hydra/Antlia is overall quite shallow,
with a peak at $v \sim 2750$\kms, which corresponds to the
extension of the Hydra/Antlia filament into the ZOA, and a broader
overdensity at about $6000$\kms, associated with the Vela overdensity
($280\deg,+6\deg$), next to some higher velocity peaks.

In the Crux region, a broad concentration of galaxies is present from
about 3500 to 8500\kms. This feature is already influenced by the GA
overdensity. It is due to a wall-like structure that seems to connect
the Norma cluster across the Centaurus-Crux, respectively the CIZA
J\,1324.7$-$5736 cluster at $(\ell,b,v) \sim (307\deg,5\deg,6200$\kms)
(Woudt 1998; Ebeling, Mullis \& Tully 2002) to the Vela overdensity at
($280\deg,+6\deg,6000$\kms) (see also Fig.~\ref{vslice} and
\ref{wedge10}).

The GA histogram, in comparison, is strongly dominated by the very
high peak associated with the Norma cluster at 4848\kms\
(Kraan-Korteweg et al.~1996; Woudt 1998) and the surrounding great
wall-like structure in which it is embedded. The peak corresponds
exactly to the predicted mean redshift of the Great Attractor.  The
Norma cluster, with a velocity dispersion of 896\kms, has been found
to have a virial mass of the same order as the Coma cluster
(Kraan-Korteweg \etal 1996; Woudt 1998).  This is confirmed
independently by the X-ray observations with ROSAT (B\"ohringer \etal
1996) which finds the Norma cluster to be the 6th brightest ROSAT
cluster in the sky.  Simulations have furthermore shown that the
well-known Coma cluster would appear the same as the Norma cluster if
Coma were located behind the Milky Way at the location of the Norma
cluster (Woudt 1998). The Norma cluster does not only seem the most
likely candidate to define the core of the GA, it also seems to have
superseded the well-known Coma cluster as being the nearest rich
cluster to the LG, and therefore is an interesting cluster on its own,
irrespective of its central position in the GA region.

A detailed dynamical analysis of the Norma cluster is in preparation
(see Woudt 1998, Woudt et al.~2000, for preliminary results), as well
as a precise determination of its distance (Woudt et al.~2005, Woudt
et al., in prep.), in order to determine whether the Norma cluster,
and therefore the GA as a whole, is at rest with respect to the CMB,
or partakes in the earlier mentioned, still controversial flow towards
to Shapley Concentration.

Besides the dominant peak due to the GA, the peak at 14\,000\kms\ in
the lower panel of Fig.~\ref{mefos} is noteworthy. It is due to the
Ara cluster (Woudt 1998; Ebeling et al.~2002) which together with the
adjacent Triangulum-Australis cluster (McHardy et al.~1981) forms a
larger overdensity referred to as a 'Greater Attractor behind the
Great Attractor' by Saunders et al.~(2000). They find a signature of
this overdensity in the reconstructed IRAS galaxy density field.

\subsection{Does the Norma Cluster Define the Core of the GA?}

The emerging optical picture of the Great Attractor so far is that of a
confluence of superclusters (the Centaurus Wall and the Norma
supercluster) with the Norma cluster being the most likely candidate
for the Great Attractor's previously unseen center. However, the
potential well of the GA might be rather shallow and extended. Seen
that the ZOA has not been completely reduced by deep 
optical surveys it is not inconceivable that further prospective galaxy
clusters might be located at the bottom of the GA's potential well at
extinction levels ${A_B} \ge 3^{\rm m}$ (Fig.~\ref{ga}). 

Detecting clusters at higher extinction levels is not straightforward.
The X-ray band is potentially an excellent window for studies of
large-scale structure in the ZOA, because the Milky Way is transparent
to the hard X-ray emission above a few keV, and because rich clusters
are strong X-ray emitters. But although dust extinction and stellar
confusion are unimportant in the X-ray band, photoelectric absorption
by the Galactic hydrogen atoms -- the X-ray absorbing equivalent
hydrogen column density -- does also limit detections close to the
Galactic Plane. A systematic X-ray search for clusters in the ZOA
($|b| \le 20\deg$) has been performed by Ebeling \etal (2002). They
found only the above-mentioned Centaurus-Crux or CIZA J\,1324.7$-$5736
cluster, as a previously unknown component, that might form part of of
the GA (see also Kocevski et al. 2004, Mullis et al. 2005), though it
is by no means as centrally located in the GA, nor as massive as the
Norma cluster.

Alternatively, a strong central radio source, such as PKS\,1610$-$608
in the Norma cluster, could also point to unidentified clusters.
Exactly such a source lies in the deepest layers of the Galactic
foreground extinction ($A_B = 12^{\rm m}$) at $(\ell, b, v) =
(309.7\deg, +1.7\deg, 3872$ km s$^{-1}$). This strong radio source was
suspected for a long time by Kraan-Korteweg \& Woudt (1999) of being
the principal member of an unknown rich galaxy cluster. An overdensity
of galaxies around this massive galaxy had indeed been seen in blind
HI-surveys, which are uneffected by extinction (see
Sect.~\ref{HI}). Moreover, dedicated infrared studies in the
surroundings of PKS\,1343$-$601 also found an excess of galaxies
(e.g. Kraan-Korteweg et al. 2005a in the $I$-band; Schr\"oder et
al. 2005 on DENIS $IHK$- images; and Nagayama et al. 2004, 2005, in
$JHK$). The data are, however, not supportive of this being a rich
massive cluster. This is consistent with the upper limit of X-ray
emission determined from ROSAT data by Ebeling et al.~2002.
 
ACO\,3627 thus remains the most likely candidate of constituting the
central density peak of the potential well of the Great Attractor
overdensity.
 
\section{2MASS Galaxies and the ZOA}\label{NIR}

Observations in the near infrared (NIR) can provide important
complementary data to other surveys. With extinction decreasing as a
function of wavelength, NIR photons are much less affected by
absorption compared to optical surveys. The \I, \J, \HH\ and $K$ band
extinction is only 45\%, 21\%, 14\% and 9\% compared to the optical
\B\ band -- hence, as the mean longitude of the surveys in the
respective wavebands increases, a progressively deeper search into
thicker obscuration layers at lower Galactic latitudes is possible.

The NIR is sensitive to early-type galaxies -- tracers of massive
groups and clusters -- which are missed in far infrared (FIR) surveys
not discussed in this paper (but see \kkl) and HI surveys
(Sect.~\ref{HI}). Moreover, recent star formation contributes only
little to the NIR flux of galaxies (in contrast to optical and FIR
emission) and therefore provides a better estimate of the stellar mass
content of galaxies.

Two systematic near infrared surveys have been performed: DENIS, the
DEep Near Infrared Southern Sky Survey, has imaged the southern sky
from $-88\deg < \delta < +2\deg$ in the \II\ (0.8\micron), \J\
(1.25\micron) and \K\ (2.15\micron) bands with magnitude completeness
limits for stars of $I=18\fm5$, $J = 16\fm5$ and $K = 14\fm0$ leading
to a prediction of the detection of 100 million stars (see
http://www-denis.iap.fr for further details). The DENIS completeness
limits (total magnitudes) for highly reliable automated galaxy
extraction away from the ZOA ($|b| > 10\deg$) was determined as $I =
16\fm5$, $J = 14\fm8$, $K_s = 12\fm0$ by Mamon (1998), leading to a
predicted extraction of roughly 250\,000 galaxies. A provisional DENIS
$I$-band catalog of galaxies with $I \le 14\fm5$ for 67\% of the
southern sky has been released by Paturel, Rousseau \& Vauglin (2003),
and over 2000 serendipitous DENIS detections of galaxies behind the
Milky Way by Vauglin et al. 2002 (see also Rousseau et al.~2000 for
the description of some noteworthy DENIS galaxies uncovered in the
ZOA). Results of pilot studies in probing the ZOA using DENIS data
concentrated on the Great Attractor region, in particular in the
surroundings of the cluster ACO\,3627 and the radio source
PKS\,1343$-$601, have been given in Schr\"oder \etal (1997, 1999, 2000,
2005) and Kraan-Korteweg \etal (1998), and \kkl.

2MASS, the 2 Micron All Sky Survey, covers the whole sky in the \J\
(1.25\micron), \HH\ (1.65\micron) and \K\ (2.15\micron) bands. Its
point source sensitivity limits are $J = 15\fm8$, $H = 15\fm1$ and $K
= 14\fm3$, whereas for galaxies and other spatially resolved objects,
the survey should be complete away from the Galactic Plane to
$J=15\fm0$, $H = 14\fm3$ and $K = 13\fm5$ over a wide range of surface
brightnesses (Jarrett et al. 2000a,b).  Source extraction resulted in
a Point Source Catalog (PSC) containing close to half a billion
objects, most of which will be Milky Way stars (next to an estimated 3
to 5 million unresolved galaxies), as well as the 2MASS Extended
Source Catalog (2MASX) which contains 1.65 million galaxies or other
extended sources.

In the following, we will discuss the effectiveness of 2MASX with
regard to Zone of Avoidance penetration and compare this to the reduced
optical ZOA including the results from the deep optical surveys. This
not only to determine the most effective method of uncovering galaxies
hidden by the Milky Way, but by studying the magnitude completeness
limits, colors and surface brightness as a function of extinction or
star density, we hope, amongst others, to also optimize redshift
follow-up observations in the ZOA.

\subsection{The 2MASX Zone of Avoidance}
The final release of the 2MASX by the Two Micron All Sky Survey Team
in 2003 (Jarrett et al. 2000b) now allows a detailed study of the
performance of 2MASS in mapping the extragalactic large-scale
structures across the ZOA, as well as compare and cross-correlate the
2MASS galaxy distribution with the deep optical ZOA catalogs (see also
Kraan-Korteweg \& Jarrett 2005).

\begin{figure}
\hfil \psfig{figure=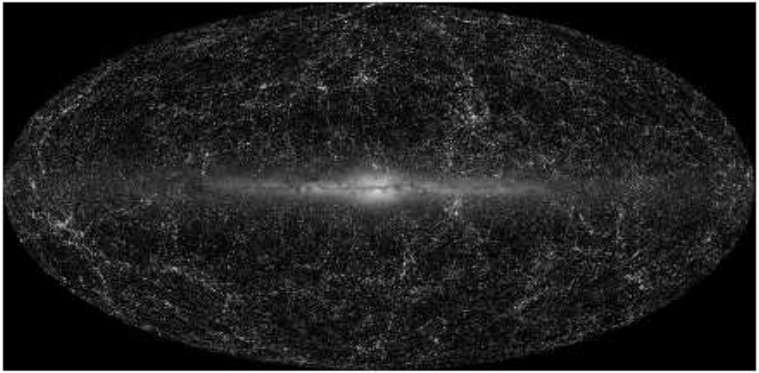,width=10cm} \hfil
\caption{Distribution of 2MASX sources with $K < 14\fm0$ in an equal
area Aitoff projection in Galactic coordinates centered on the
Galactic Bulge. Stars from the PSC area also displayed. Figure adopted
from Jarrett 2004.}
\label{2MASS}
\end{figure}

Fig.~\ref{2MASS} (adopted from Jarrett 2004) shows an Aitoff
projection in Galactic coordinates centered on the Galactic Bulge of
all the 1.65 million resolved MASX sources with magnitudes brighter
than $K < 14\fm0$, in addition to the nearly 0.5 billion Milky Way
stars. A description of the large-scale structure of these galaxies
based on redshifts estimated from their NIR colors is given in Jarrett
2004. Here we will concentrate on the penetration of the ZOA.

Compared to the optical, the 2MASX sources provide a much deeper and
uniform view of the whole extragalactic sky.  The galaxy distribution
can be traced without hardly any hindrance in the Galactic
Anticenter. However, the wider Galactic Bulge region -- represented
here by the half billion Galactic stars from the PSC -- continues to
hide a non-negligible part of the extragalactic sky.  Despite the fact
that NIR surveys should in principle be able to uncover galaxies to
extinction levels of about $A_B = 10^{\rm m}$ compared to $3^{\rm m}$
in the optical, the NIR ZOA does not appear narrower here compared to
the reduced optical ZOA which on average has an approximate width of
$b \la \pm5\deg$ around the Galactic Plane (see Fig.~4 in \kkl\ which
shows a whole-sky distribution of galaxies with $D \ge 1\farcm3$ that
has been complemented with all published ZOA galaxies that also meet
that criterium).

Thus, there clearly also exists a NIR ZOA, but its form is quite
distinct from the optical one. To understand the differences between
these two ZOAs we compared in detail the results from our optical
survey in the Scorpius region with 2MASX detections. The Scorpius
region lies to the left of the GA region (most left search area in
Fig.~\ref{WKK}) and to the right of the Galactic Center, where
confusion from the foreground Milky Way is extreme.

\begin{figure}[t!]
\hfil \psfig{figure=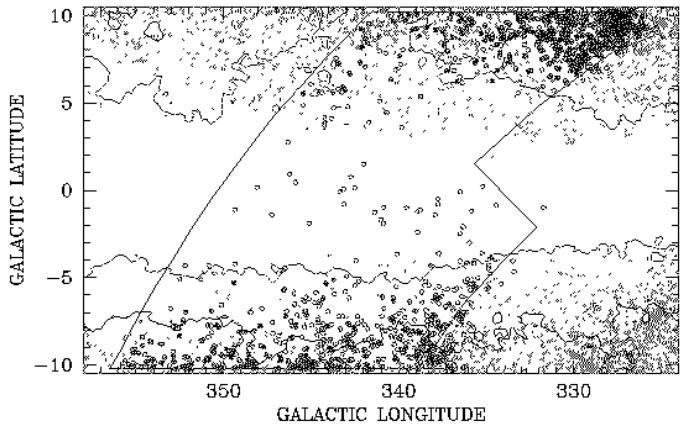,width=9cm} \hfil
\caption{Distribution of optically detected galaxies (circles) with 
$D > 12\arcsec$ in the Scorpius regions,
and 2MASS galaxies (small dots) with $K \le 14\fm0$ 
and $(J-H) > 0\fm0$ in the Scorpius search region and
surroundings. The contours mark extinction levels of $A_B= 1\fm0$ and
$3\fm0$.}
\label{sco}
\end{figure}

The optical and 2MASS galaxy distributions are displayed jointly in
Fig.~\ref{sco}. The large circles represent optically detected
galaxies in Scorpius (Fairall \& Kraan-Korteweg 2000, 2005), the small
dots 2MASX objects with $K \le 14\fm0$. It should be noted that
extremely blue objects ($(J-H) < 0\fm0$) are excluded, as well as
2MASX sources that were rejected as likely galaxy candidates upon
visual examination by Jarrett (priv. comm.). These generally are
Galactic objects, such as HII regions and Planetary Nebulae, the prime
contaminant at $|b| < 2\deg$ (optical catalogs also contain a small
fraction of them). The 1\fm0 and 3\fm0 optical extinction contours are
also drawn.

Fig.~\ref{sco} confirms that even this close to the Galactic Bulge,
where star densities are high, deep optical searches are fairly
faithful tracers of the galaxy distribution to $A_B \sim 3^{\rm m}$,
with only a few -- mostly uncertain -- galaxy candidates peaking
through higher dust levels. Though the extinction in the $K$-band is
only 9\% of that in the $B$-band, only about one-third of the
optically identified galaxies in the Scorpius region have a
counterpart in 2MASS.  Although the 2MASX sources seem to probe deeper
into the Milky Way above the Galactic Plane, this trend is not seen at
negative latitudes. There, clearly optical galaxies dominate and they
also probe the galaxy distribution deeper into the plane.

A direct correlation between dust absorption and 2MASX source density
is not seen here. NIR surveys become progressively less succesful
compared to optical surveys when approaching the Galactic Center.  In
fact, this progressive loss of 2MASX galaxies is already noticeable in
the fraction of optical galaxies that have counterparts in the 2MASX
catalog.  This fraction is 47\% in the Hydra/Antlia region ($\ell
\approx 280\deg$; see Fig.~\ref{WKK} for orientation), and decreases
to 39\% in the Crux region ($\ell \approx 310\deg$), 37\% in the GA
region ($\ell \approx 325\deg$), 33\% in the Scorpius region ($\ell
\approx 340\deg$), and to a mere 16\% in the by Wakamatsu et al. (2000,
2005) explored Ophiuchus cluster region ($\ell=0\fdg5,
b=+9\fdg5$). 

It should be maintained though, that there also are 2MASX galaxies in
the optically surveyed regions that have no optical counterpart.  This
fraction increases inversely, i.e. the farther away from the Galactic
Bulge the higher this fraction. Not unsurprisingly, the majority of
ZOA galaxies that are seen in 2MASS galaxies but not in the optical are
on average quite red and faint (mostly with $K \ga 12^{\rm m}$). This
is partly a ZOA effect with NIR surveys finding red galaxies more
readily at higher absorption level. But it is also due to the inherent
characteristics of the two surveys. Whereas the NIR is better at
detecting old galaxies and ellipticals, the optical surveys --
although susceptible to all galaxy types -- are best at finding
spirals and late-type galaxies (especially low surface-brightness
galaxies and dwarfs).  The surveys are in fact complementary.

\begin{figure}[p]
\hfil \psfig{figure=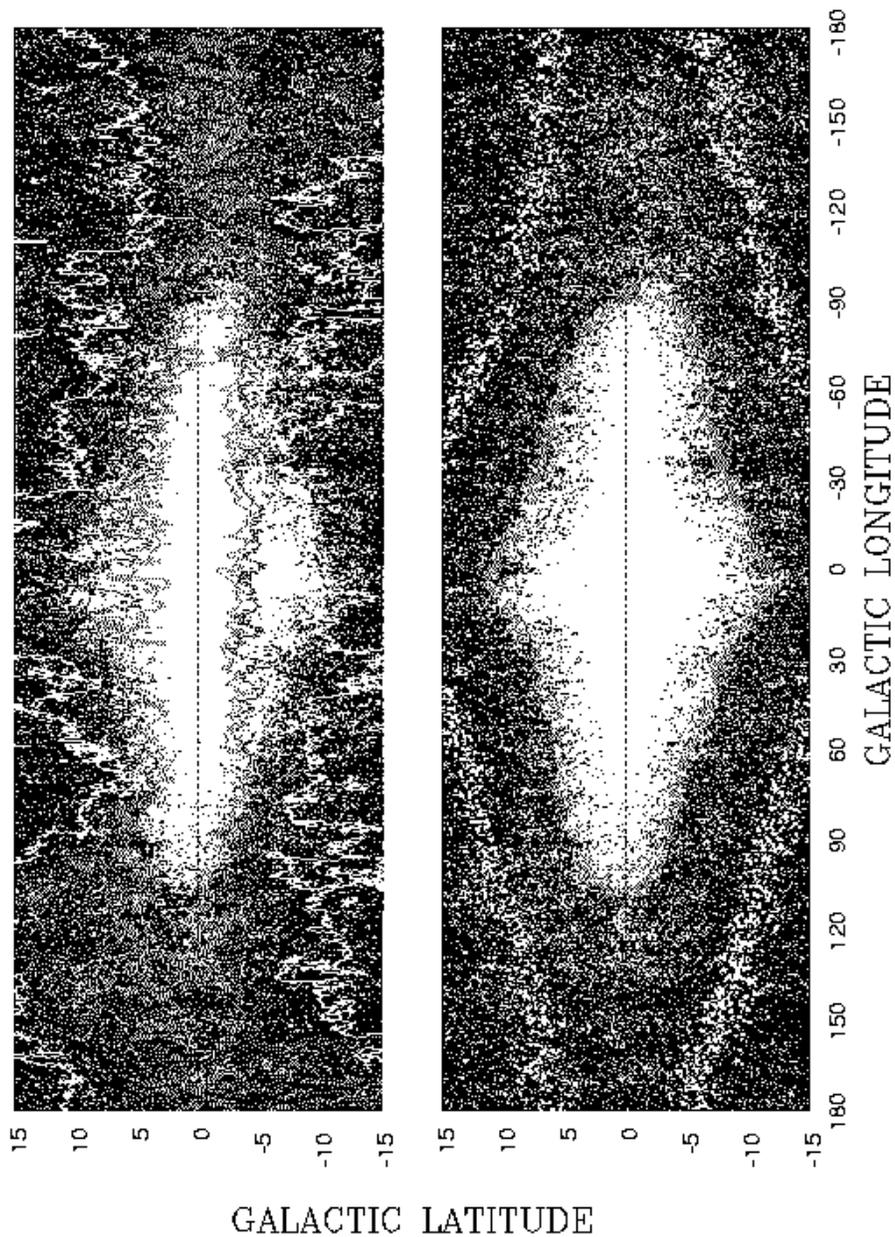,angle=0,width=11.8cm} \hfil
\caption{Distribution of 2MASX sources with $K \le 14\fm0$ along the
Galactic equator within $b \le \pm 15\deg$.
In the left panel, DIRBE/IRAS extinction contours of $A_B=1\fm0,
3\fm0$ and $5\fm0$ are superimposed. In the right panel, galaxies
found in regions of star densities of log$N = 3.50$, 3.75, and 4.00
per square degree ($K \le 14\fm0$) are enhanced.}
\label{2M_ZOA}
\end{figure}

The success rate of galaxy identification in the NIR actually depends
much more strongly on {\sl star density} than dust extinction. This
effect is corroborated by Fig.~\ref{2M_ZOA}, which shows 2MASX sources
with $K \le 14\fm0$ within $\pm 15 \deg$ of the Galactic equator. In
the left panel, DIRBE/IRAS extinction contours of $A_B = 1\fm0, 3\fm0$
and $5\fm0$ are superimposed. The right panel emphasizes the locations
of 2MASX sources in regions where the density of stars in the PSC with
$K \le 14\fm0$ per square degree is log$N$ = 3.50, 3.75, and 4.00.

An examination of the dust extinction contours with the area in which
2MASX does not find extended sources does not suggest a correlation
between them.  In the Galactic Anticenter, roughly defined here as
$\ell \sim 180\deg \pm 90\deg$, 2MASX objects seem to cross the Plane
without any hindrance. For this half of the ZOA, plots of NIR
magnitude or diameter versus extinction (not shown here) confirm that
galaxies can be easily identified up to extinction levels equivalent
to $A_B \ga 10^{\rm m}$, and extinction-corrected $K^o$-band
magnitudes versus extinction diagrams imply that 2MASS remains quite
complete up to $K^o \la 13\fm0$ for $A_B \sim 10\fm5$.

This is not at all true for the wider Galactic Bulge region ($\ell
\sim 0\deg \pm 90\deg$), where 2MASX detects objects to lower
extinction levels only and where the completeness limit is at least
one magnitude lower compared to the Anticenter. And although Galactic
dust does reduce the completeness limit of 2MASX sources at low
latitudes -- although to a much lower extent than in the optical --
the origin of the NIR ZOA is mainly due to source confusion in regions
of high {\sl star density}.

This region in which NIR surveys fail completely has a very
well-defined shape. It is traced by the star density
isopleth (stars per square degree with $K < 14\fm0$) of log $(N) =
4.00$ , i.e. the innermost contour in the right panel of
Fig.~\ref{2M_ZOA}. At this level, the completeness has 
already dropped significantly. Above this limit, the point 
sources are packed so densely that extended sources can not be
extracted anymore. The hoped-for improvement of uncovering the galaxy
overdensity with NIR surveys to lower latitudes compared to the
optical, as for instance in the Great Attractor region (compare
Fig.~\ref{WKK} or \ref{ga} to Fig.~\ref{2MASS}), has not been achieved.

It should be maintained, however, that for the ZOA away from the Bulge
-- as well as for the rest of the sky -- 2MASS as a homogeneous
whole-sky survey obviously is far superior. An optimal approach in
revealing the galaxy distribution behind the Milky Way will actually
result from a combination of deep optical and near-infrared
surveys. Not only because of the different susceptibility to
morphology, but because the former are sensitive to galaxies located
behind regions of high source confusion, and the latter to galaxies
located behind thick dust walls. 

A reduction of the ZOA common to both the optical and NIR might be
achieved by a combination of a deep $R$-band survey with a spatially
higher resolved NIR survey (higher than the current NIR surveys 2MASS
and DENIS) which should diminish the source confusion in high star
density regions.

\subsection{The Shape of the NIR ZOA}

For the discussion in this section on the shape of the NIR ZOA, we
define it more or less ad hoc as the nearly completely empty region as
outlined by the isopleth given with the 2MASS PSC star density of
log$N = 4.00$ per square degree for stars with $K \le 14\fm0$ (inner
contour of left panel of Fig.~\ref{2M_ZOA}). A closer look at the NIR
ZOA reveals some interesting asymmetries that are independent of
extragalactic large-scale structure but are actually due to Galactic
structure.

The bulge and disk as outlined by the lack of galaxies seem to be
inclined with respect to the Galactic equator. Its mean latitude is
offset to positive latitudes for $\ell \sim 90\deg$, and to negative
latitudes for $\ell \sim 270\deg$ ($-90\deg$).  This lopsidedness has
been known for a long time. It was first established from the Galactic
hydrogen column densities by Kerr \& Westerhout in 1965 for longitudes
between $\ell = 220\deg - 330\deg$. The same inclination is evidenced
also in the dust contours (see left panel of Fig.~\ref{2M_ZOA}).

Another asymmetry becomes obvious when regarding the width of the NIR
ZOA with respect to the Galactic Center. Whereas the NIR ZOA can be
followed to about $\ell \sim 120\deg$ on the one side, it stretches
over 'only' $90\deg$ on the opposite side of the Galactic Center (to
$\ell \sim 270\deg$). The explanation for this asymmetry lies in the
relative location of the Sun with respect to the Galactic spiral
arms. When looking towards the local Orion arm, our line of sight is
hit directly with a high density of nearby stars, blocking a larger
fraction of the extragalactic sky from our view, whereas the opposite
line of sight is nearly free of stars for quite a distance until it
hits the more distant spiral arm, allowing us to identify
galaxies more easily between the fainter and smaller stars of this
star population.

Furthermore, when regarding the location where the Galactic Bulge is
highest, one notes that it does not peak at $\ell = 0\deg$ as
expected, but is centered on $\ell \sim +5\deg$. This offset probably
is due to the bar of our Galaxy. Its near side points towards us
(positive longitudes), reducing our view of the extragalactic sky
stronger compared to our line of sight towards the far side of the
Galactic bar.

As these asymmetries have more to do with Galactic structure than
extragalactic structure, one might be tempted to ignore them. However,
these asymmetries in the distribution of galaxies should be taken into
account when using the 2MASX catalog -- or redshift surveys based on
2MASX subsamples -- for dipole determinations. These structures might 
have a significant effect on the results if not properly corrected for.

\subsection{A Redshift Zone of Avoidance}\label{zZOA}

\begin{figure}
\hfil \psfig{figure=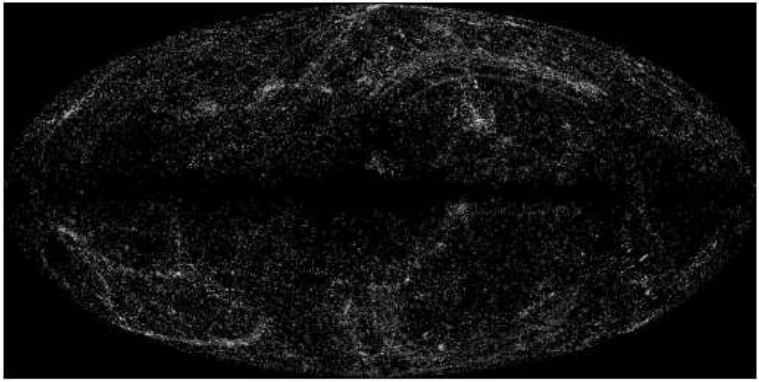,width=10cm} \hfil
\caption{Distribution of 2MASX sources in an equal area Aitoff projection
in Galactic coordinates that have redshifts listed in NED
(Jarrett 2004, priv. comm.).}
\label{2MASSz}
\end{figure}

A reduction of the ZOA on the sky does not imply at all that a similar
reduction can also be attained in redshift space. This is seen most
clearly in Fig.~\ref{2MASSz} which displays the distribution of 2MASX
galaxies that have a redshift listed in NED, the NASA/IPAC
Extragalactic Database (Jarrett 2004, priv. comm.). The ZOA in this
figure is quite distinct from the previously defined NIR ZOA. Its form
is actually much more reminiscent of the optical ZOA delimited by the
extinction contour of $A_B = 3\fm0$ (see Fig.~4 in \kkl). Hardly any
redshifts are available for latitudes of $b \la 5\deg$.

One might argue that this is an artifact because it is based mainly on
optically selected targets. 2MASX has been released only fairly
recently and systematic follow-up redshifts observations of the newly
uncovered galaxies at low Galactic latitudes, in particular in the
Galactic Anticenter ZOA half, have not yet been made and/or
published. This is however not entirely true. In whatever kind of
waveband a ZOA galaxy is identified, it remains inherently difficult
to obtain a reliable optical redshift when the extinction in the
optical towards this galaxy exceeds 3 magnitudes, i.e. the delimiting
factor of optical surveys in general.

\begin{figure}
\hfil \psfig{figure=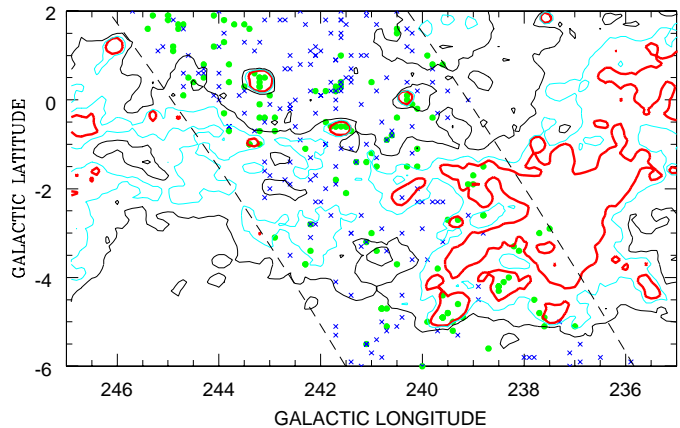,width=9cm} \hfil
\caption{Distribution of 2MASS galaxies observed with 6dF in a strip
crossing the Galactic Plane in the Puppis region.  Circles represent
galaxies with redshifts, crosses those without a reliable redshift
determination. The contours indicate an optical extinction of
$A_B=1\fm0, 2\fm0$ and $3\fm0$ (thick contour).}
\label{6df}
\end{figure}

This is seen quite clearly in Fig.~\ref{6df} which shows 2MASS
galaxies in Puppis -- a filament is crossing the Plane there (see
Fig.~\ref{2MASS}) -- on a sequence of 6-degree fields (6dF) centered
on Dec $= -25\deg$. This strip has been observed with the multifibre
spectroscope at the UK Schmidt Telescope as part of a pilot project
aimed at extending the 6dF Galaxy Survey towards lower latitudes. It
now is restricted to the southern sky with $|b| \ge 10\deg$ (see
http://www.mso.anu.edu.au/6dFGS for further details).

The crosses mark galaxies for which a reliable redshift could be
measured, the filled dots galaxies for which this could not be
realized. At first glance, the plot seems to indicate a fantastic
success rate for obtaining redshifts all the way across the Milky Way.
However, it should be noted that the concerned ZOA strip lies in a
region renowned for its low dust content (it hardly exceeds 3
magnitudes), and a careful inspection indicates that redshifts have
generally not been obtained for galaxies that lie in pockets where
the extinction is higher than $A_B \ga 3\fm0$ (thick
contour).  So even when galaxies are identifiable deep in the Plane,
reducing the {\sl redshift ZOA} will remain hard,  and an
optimization of targeted ZOA galaxies is crucial for redshift
follow-ups.

\section{Dedicated HI Galaxy Searches in the ZOA}\label{HI}

Because the Galaxy is fully  transparent to the 21cm line radiation of
neutral hydrogen,  HI-rich galaxies can  readily be found  through the
detection  of their  redshifted 21cm  emission in  the regions  of the
highest  obscuration  and infrared  confusion.  Furthermore, with  the
detection of an  HI signal, the redshift and  rotational properties of
an external  galaxy are immediately known, providing  insight not only
on its location in redshift space but also on the intrinsic properties
of  such obscured  galaxies. This  makes systematic  blind  HI surveys
powerful tools in mapping large-scale structures behind the Milky Way.

Early-type galaxies -- tracers of massive groups and clusters
-- are gas-poor and will, however, not be identified in these
surveys. Furthermore, low-velocity extragalactic sources that fall
within the velocity range of the strong emission of the Galactic gas ($v
\la \pm 250$\kms) will be missed. Galaxies that lie close in position 
to radio continuum sources may also be missed because of the baseline 
ripples they produce over the whole observed requency range.
 
Two systematic blind HI searches for galaxies behind the Milky Way
have been made. The first used the 25~m Dwingeloo radio to survey the
whole northern Galactic Plane for galaxies out to 4000~\kms\ with a
sensitivity of $rms$ of 40~mJy for a 1~hr integration (see \kkl\ for a
summary of the results). A more sensitive survey ($rms$ of typically 6
mJy beam$^{-1}$), probing a considerably larger volume (out to
12\,700~\kms), has been performed with the Multibeam Receiver at the
Parkes 64\,m radio telescope in the southern sky. In the following,
the most recent results of this survey are given.

\subsection{The Parkes Multibeam HI ZOA Survey}

The Multibeam receiver at the 64\,m Parkes telescope was specifically
constructed to efficiently search for galaxies of low optical surface
brightness, or galaxies at high optical extinction, over large areas of
the sky. It has 13 beams, each with a beamwidth of $14\farcm4$,
arranged in a hexagonal grid in the focal plane array (Staveley-Smith
\etal 1996) which allows -- with its large footprint of 2\fdg5 on the
sky -- rapid sampling of large areas.
 
In March 1997, this instrument was mounted on the telescope and
various surveys were started, one being a systematic blind HI survey
within $b < \pm 5\deg$ of the ZOA. The observations were performed in
scanning mode.  Fields of length $\Delta\ell=8\deg$ centered on the
Galactic Plane were surveyed along constant Galactic latitudes where
each scan was offset by 35\arcmin\ in latitude until the final width
of $\Delta b = \pm 5\deg$ had been attained (17 passages back and
forth). The final goal was 25 repetitions per field. With an effective
integration time of 25 min/beam, a 3\,$\sigma$ detection limit of
25\,mJy was obtained. The correlator bandwidth of 64~MHz was set to
cover a velocity range of $-1200 \la v \la 12700$~\kms. The survey
therewith is sensitive to normal spiral galaxies well beyond the Great
Attractor region (e.g. $5 \cdot 10^9$ M$_{\odot}$ at 60 Mpc for a
galaxy with a linewidth of 200\kms), next to the lowest mass dwarf
galaxies in the local neighborhood ($10^6 - 10^7$ M$_{\odot}$), or
extremely massive galaxies beyond 10\,000\kms\ such as the
extraordinarily massive galaxy HIZOA\,J0836-43 with a HI mass of $7
\cdot 10^{10}$ M$_{\odot}$ found in one of the ZOA data cubes
(Kraan-Korteweg \etal 2005b; Donley et al. in prep.).
 
The data are in the form of three-dimensional data cubes
(position-position-velocity, with pixel and beam sizes of $4\arcmin
\times 4\arcmin$, and $15\farcm5$, respectively). Experimentation with
automatic galaxy detection algorithms indicated that visual inspection
of the data cubes is more efficient for the ZOA, where the noise due
to continuum sources and Galactic HI is high and variable. The ZOA
cubes were inspected by at least two, sometimes three, individual
researchers, with the subsequent neutral evaluation of inconsistent
cases in the detection lists by a third party.

An first analysis covering the {\sl southern} Milky Way ($212\deg \le
\ell \le 36\deg$) based on 2 out of the foreseen 25 passages (the HI
ZOA Shallow Survey; henceforth HIZSS) with and $rms$ noise of 13~mJy
beam$^{-1}$ led to the discovery of 110 galaxies, two thirds of which
were previously unknown (Henning et al.~2000). A final catalog of the
23 central cubes of the full-sensitivity survey is in preparation
(Henning et al., in prep.).

The data and plots of the full sensitivity survey are presented in the
next section. They are based on a provisional version of this catalog
which might still contain a few galaxy candidates that will be rejected
for inclusion in the final catalog. They furthermore include
detections from an extension to the north (Dec$ > 0\deg$), which was
done at a later stage, resulting in 2 further cubes on both sides of the
southern ZOA (Donley et al. 2005). The data set regarded here thus
consists of 27 data cubes that cover the ZOA between $196\deg \le \ell
\le 52\deg$ for $|b| \le 5\deg$. A total of slightly over one thousand
galaxies were identified in these data cubes.

\subsection{The Detected Galaxies}

\begin{figure}[h] 
\hfil \psfig{figure=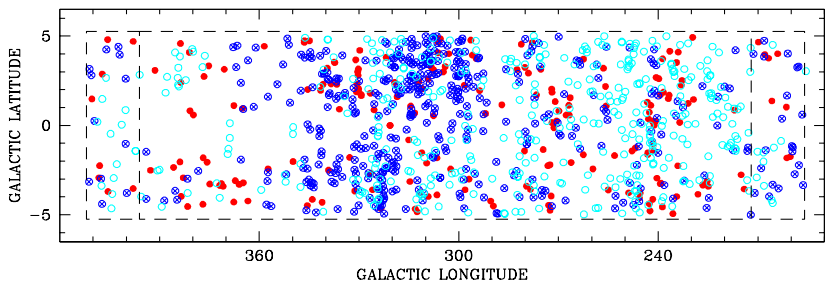,width=11cm} \hfil
\caption{Distribution in Galactic coordinates of the galaxies
detected in the deep HI ZOA survey. Open circles: $v_{\rm hel} <
3500$; circled crosses: $3500 < v_{\rm hel} < 6500$; filled circles:
$v_{\rm hel} > 9500$\kms.}
\label{sky}
\end{figure}

Figure~\ref{sky} displays the distribution along the Milky Way of the
in HI detected galaxies. An inspection of this distribution shows that
the HI survey nearly fully penetrates the ZOA with hardly any
dependence on Galactic latitude.  

\begin{figure}[t]
\hfil \psfig{figure=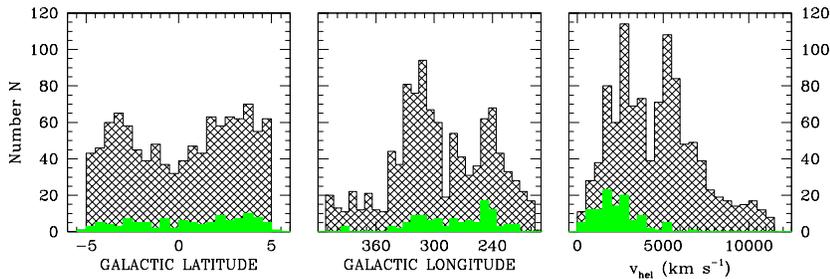,width=11cm} \hfil
\caption{Distributions as a function of the Galactic latitude,
longitude and the heliocentric velocity of the in HI detected
galaxies.  The lower histograms represent the results from the HIZSS.}
\label{hist}
\end{figure}

This is confirmed by the left panel in Fig.~\ref{hist}, which shows
the detection rate as a function of Galactic latitude. The small dip
in the detection rate between $-2\deg \la b \la +1\deg$ stems mainly
from the Galactic Bulge region and to a lesser extent from the GA
region ($\ell \approx 300\deg-340\deg$). The former is due to the high
number of continuum sources at low latitudes in the Galactic Bulge
region. In the GA region the gap is possibly related to the high
galaxian density for the on average higher velocity, hence fainter,
galaxies. There, a moderate number of continuum sources may already
result in a detection-loss. This explanation is supported by the fact
that this dip is not noticeable in the shallower HIZSS data (lower
histogram).

A much stronger variation is apparent in the number density as a
function of Galactic longitude (see also middle panel of
Fig.~\ref{hist}). This can be explained entirely with large-scale
structures such as the nearby (and therefore prominent in HIZSS) Puppis
filament ($\ell \approx 240\deg$), the Hydra-Antlia filament ($\ell
\approx 280\deg$), the very dense GA region ($\ell \approx
300-340\deg$), followed by an underdense region ($52\deg \ga \ell \ga
350\deg$) due to the Local and Sagittarius Void.

These large-scale structures have left their imprint also on the
velocity diagram (right panel of Fig.~\ref{hist}), which shows two
conspicuous broad peaks. The low-velocity one is due to a blend of
various structures in or crossing the Galactic Plane while the second
around 5000\kms\ clearly is due to the GA overdensity (see also
Fig.~\ref{wedge5} $-$ \ref{wedge10}).  The velocity histogram moreover
shows that galaxies are found all the way out to the velocity limit of
the survey of $\sim 12\,000$\kms, hence probe the galaxy distribution
considerably deeper than either the shallow ZOA survey HIZSS (lower
histogram), or the southern sky HI surveys also made with the Multibeam
instrument, the HI Bright Galaxy Catalog (BGC; Koribalski et al.\
2003) and the HI Parkes All Sky Survey (HIPASS; Meyer et al.~2004).

\subsection{Uncovered Large-Scale Structures in the GA}

\begin{figure}[t] 
\hfil \psfig{figure=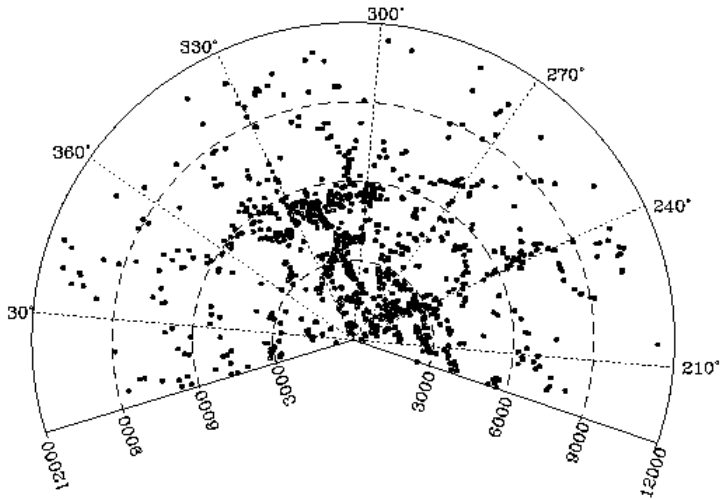,width=10cm} \hfil
\caption{Galactic latitude slice with $|b| \le 5\deg$ out to
12000\kms\ of the slightly over 1000 in HI detected galaxies. Circles
mark intervals of 3000\kms.}
\label{wedge5}
\end{figure}

Figure~\ref{wedge5} shows a Galactic latitude slice with $|b| \le
5\deg$ out to 12\,000\kms\ of the galaxies detected in the deep Parkes
HI ZOA survey (henceforth HIZOA) for the longitude range $196\deg \le
\ell \le 52\deg$. The clear conclusion when inspecting this figure is
that the HI survey really permits the tracing of large-scale structures in
the most opaque part of the ZOA; and this in a homogenous way,
unbiased by the clumpiness of the foreground dust contamination.

\begin{figure}[t!]
\hfil \psfig{figure=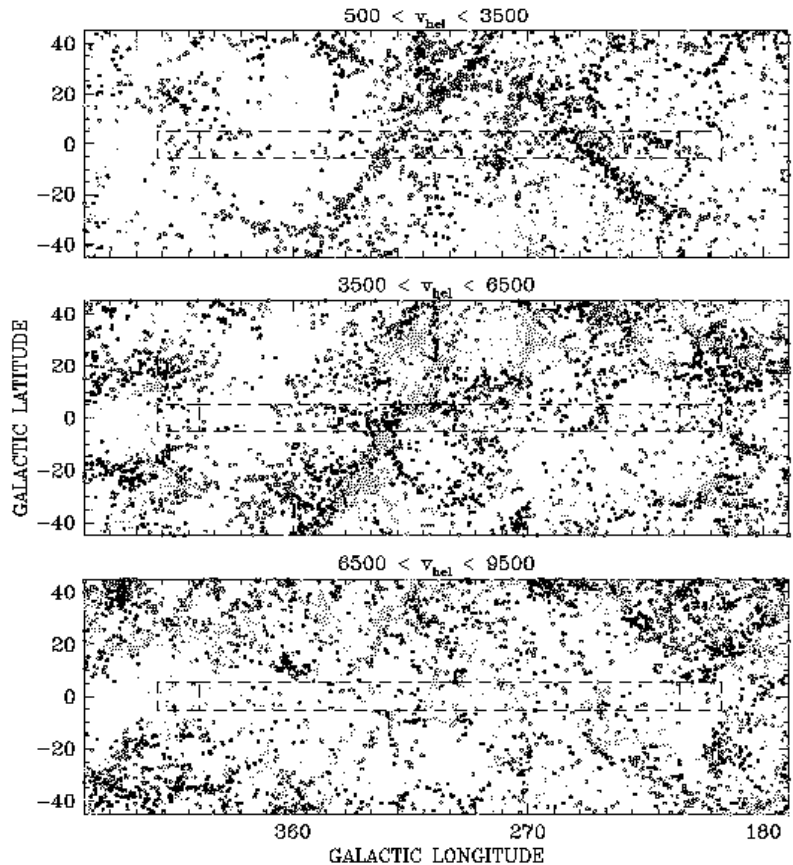,width=10.5cm} \hfil
\caption{Sky projections of three redshift slices of depth 
$\Delta v = 3000$\kms\ showing the HIZOA data in combination with 
data from LEDA. The HIZOA survey area is outlined.}
\label{vslice}
\end{figure}

In the following, some of the most interesting features revealed in
Fig.~\ref{wedge5} will be discussed. It is suggested to simultaneously
consult Fig.~\ref{vslice}, which shows the HIZOA data with data
extracted from LEDA surrounding the ZOA in sky projections for three
velocity shells of thickness 3000\kms. This helps to show the newly
discovered features in context to known structures. Viewing the
distribution of galaxies within $\Delta b \le 5\deg$ in this figure
illustrates quite clearly how the HI survey has managed to fill in
that part of the ZOA, tracing various contiguous structures across the
plane of the Milky Way.

The most prominent large-scale structure in Fig.~\ref{wedge5}
certainly is the Norma Supercluster which seems to stretch from
$360\deg$ to $290\deg$ in this plot, lying always just below the
6000\kms\ circle, with a weakly visible extension towards Vela ($\sim
270\deg$). The latter is more pronounced at higher latitudes (see
panel 2 in Fig.~\ref{vslice}, and Fig.~\ref{wedge10}). This wall-like
feature seems to be formed of various agglomerations. The first one
around $340\deg$ is seen for the first time with new HI data. Because
of the high extinction there, it cannot be assessed whether this
overdensity continues for $|b| > 5\deg$. The clump at $325\deg$ is due
to the outer boundaries at lower latitudes side of the Norma cluster
A3627 ($\ell,b,v = 325\deg,-7\deg,4880$\kms; Kraan-Korteweg et
al.~1996). The next two are previously unrecognized due to groups (or
small clusters) at $310\deg$ and $300\deg$, both at $|b| \sim
+4\deg$. They are very distinct in the middle panel of
Fig.~\ref{vslice}. Between these two clusters at slightly higher
latitude we see a small finger of God, which belongs to the
Centaurus-Crux/CIZA J\,1324.7$-$5736 cluster.

This is not the only overdensity in the GA region. A significant
agglomeration of galaxies is evident closer by at $(\ell,v) =
(311\deg,3900$\kms) with two filaments merging into it. Although no
finger of God is visible (never very notable in HI redshift slices),
this concentration forms part of the previously discussed galaxy
concentration around the strong radio source PKS\,1343$-$601 which is
consistent with an intermediate size cluster residing there.

Next to the Norma cluster, only the Centaurus-Crux cluster has an
appreciable X-ray emission (Ebeling et al. 2002; Mullis et al. 2005).
It seems therefore unlikely that any of the newly apparent galaxy
concentrations are a signature of a further massive cluster that
conforms part of the GA overdensity. But it should be kept in mind
that X-ray photons are subjected to photoelectric absorption by the
Galactic hydrogen atoms -- the X-ray absorbing equivalent hydrogen
column density -- which does limit detections close to the Galactic
Plane. This effect is particularly severe for the softest X-ray
emission, as observed by ROSAT ($0.1$-$2.4$~keV), and is seen in the
CIZA cluster distribution (CIZA standing for Clusters in the Zone of
Avoidance, a systematic search for X-ray clusters with Galactic
latitiudes $|b| \le 20\deg$). Hardly any clusters are found for
Galactic HI column density over $N_{\rm HI} > 5 \times 10^{21}$
cm$^{-2}$, which creates an X-ray ZOA of about $\Delta b \la 5\deg$ on
average (see Fig.~14 in \kkl; Fig.~7 in Ebeling et al. 2002).

Although the overdensity in the GA region looks quite impressive here,
a preliminary quantitative analysis of the 4 cubes covering $300\deg
\le \ell \le 332\deg$ by Staveley-Smith et al. (2000) find a mass excess
of 'only' $\sim 2 \cdot 10^{15}\Omega_0 {\rm M_\odot}$ over the
background, thus considerably lower than the predictions from the
infall pattern. Then again, the signatures of overdensities and
clusters are overall much shallower in HI surveys compared to, e.g.,
optical surveys (e.g. Fig.~22 versus Fig.~23 in Koribalski et
al. 2004).

To the left of the GA overdensity a few other features are worthwile
describing. An underdense region comprised of the Local Void and the
Sagittarius Void is seen around $\ell=360\deg$ at central velocities
of $\sim 1500$ and 4500\kms\ (see also the first two panels of
Fig.~\ref{vslice}). Except for the tiny group of galaxies at about
($350\deg,3000$\kms), the distribution here and in Fig.~\ref{wedge5}
suggest one big void rather than two seperate ones. In contrast, the
righthand side of Fig.~\ref{vslice} is quite crowded: the Puppis
region ($\ell \sim 240\deg$) with its two nearby groups (800 and
1500\kms) followed by the Hydra Wall at about 3000\kms\ that extends
from the Monocerus group ($210\deg$) to the concentration at
$280\deg$. The latter is not the signature of a group but due to a
filament emerging out of the Antlia cluster ($273\deg,19\deg$; see
Fig.~\ref{vslice}).

With this systematic HI survey, we could map for the first time
large-scale structures without any hindrance across the Milky Way
(Figs.~\ref{wedge5} and \ref{vslice}). It is the only approach that
easily uncovers galaxies in the ZOA - and records their redshift. For
this reason we are currently extending the Parkes HI ZOA survey to
higher Galactic latitudes in the Galactic Bulge region ($332\deg <
\ell < 36\deg$) where the optical and NIR ZOAs are wider and knowledge
about the structures very poor. They will improve the knowledge on the
borders of the Local and Sagittarius Void, as well as the Ophiuchus
cluster studied optically by Wakamatsu et al. (2000; 2005).

\section{Discussion}\label{Sum}

In the last decade, enormous progress has been made in unveiling the
extragalactic sky behind the Milky Way. At optical wavebands, the
entire ZOA has been systematically surveyed, reducing the optical ZOA
by about a factor of $2 - 2.5$, i.e. from ${A_B} = 1\fm0$ to ${A_B} =
3\fm0$. Its average width is about $\pm5\deg$, except in the
low-extinction Puppis area, where galaxies have been found at all
latitudes, in contrast to the Galactic Bulge region where the $3\fm0$
contour rises to higher latitudes for positive latitudes.

2MASS, as a homogeneous NIR survey, obviously is far superior to
optical surveys, particularly considering that the optical ZOA surveys
were not only performed on different plate material, but also by
different searchers using different search techniques. Nevertheless,
in the regions of highest star densities (over 10\,000 stars with $K
\le 14^{\rm m}$ per square degree), i.e. around the Galactic Bulge
($\ell\la \pm90\deg$), the identification of galaxies fails for
latitudes between $\pm5\deg$ up to $\pm10\deg$ (see innermost contour
in right panel of Fig.~\ref{2M_ZOA}) and the hoped-for improvement of
uncovering further galaxy overdensities in, for instance, the Great
Attractor, could not be realized.

Even when a galaxy can be identified at high extinction levels, such
as is possible in the NIR and FIR, this will, however, not
automatically reduce the ZOA in redshift space. As discussed in
Sect.~\ref{zZOA} (see Fig.~\ref{2MASSz}) it remains nearly impossible
to obtain optical redshifts for galaxies at extinction levels $A_B \ga
3^{\rm m}$. At these levels only HI observations prevail -- if the
galaxies are gas-rich and not too distant. But even with a HI
detection, cross-identification with its optical, 2MASS and/or IRAS
counterpart (Donley et al. 2005) often remains ambiguous because of
positional uncertainty due to the large beams of single-dish
radio telescopes.

\begin{figure}[t] 
\hfil \psfig{figure=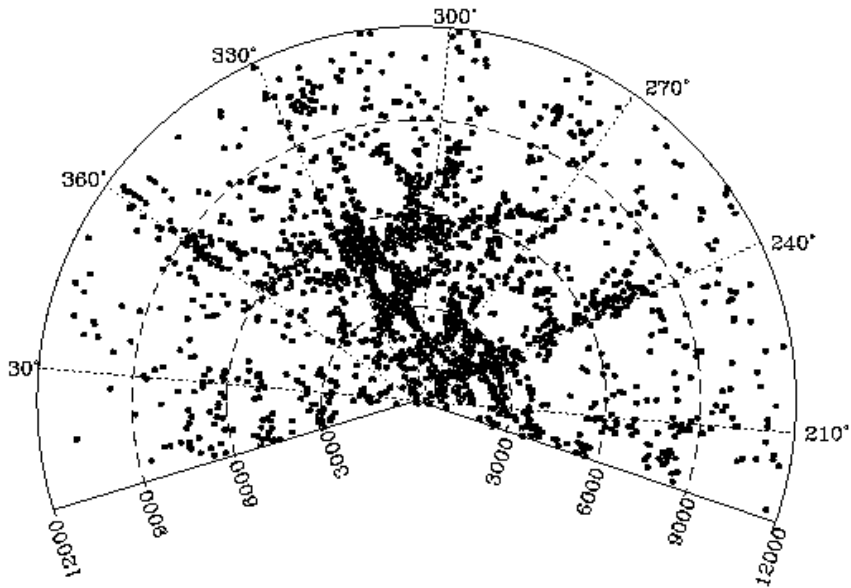,width=11.8cm} \hfil
\caption{Galactic latitude slice within $|b| \le 10\deg$ out to
12000\kms\ with all the galaxies extracted from LEDA, including the HI
Multibeam data displayed in Fig.~\ref{wedge5}.}
\label{wedge10}
\end{figure}

As seen in this paper, the mapping of the galaxy distribution behind
the Milky Way requires considerable efforts. Still, combining data
obtained from the various multi-wavelength approaches will reveal this
hidden part of the Universe, as clearly illustrated with
Fig.~\ref{wedge10} for the Great Attractor region -- though a
quantification of the structures will remain difficult due to the
different biases and selection effects in the different methods.

Figure~\ref{wedge10} shows a redshift slice of width $|b| \le 10\deg$
out to 12\,000\kms\ compared to the $\pm 5\deg$ of the HI data
alone. All galaxies with a redshift in the LEDA data base are included
next to the data from the HIZOA. Note that this is not a homogenous
data set. It clearly is deeper sampled from $270\deg \la \ell \la
340\deg$ because of the intensive redshift follow-up programs of the
ZOA catalogs by Kraan-Korteweg, Woudt and collaborators in that
longitude range (see Sect.~\ref{vel}).

Before the ZOA research programs, this slice only had a few points in
them -- mainly in the low extinction Puppis area ($\ell\sim 240\deg$)
-- and certainly did not allow any reliable description of large-scale
structures. It now has been filled to a depth comparable to unobscured
regions in the sky and the above diagram reveals various clusters,
filamentary and wall-like structures, next to some sharply outlined
voids which finally allow a fairly profound glimpse at the previously
hidden core of the Great Attractor.  Fig.~\ref{wedge10} clearly shows
the prominence of the Norma cluster ($\ell = 325\deg$) as well as its
central location in the great-wall like structure that can be followed
from $270\deg$ to $360\deg$ within the redshift range
4000$-$6000\kms. In front of this wall, we see further filaments that
merge in the galaxy concentration around PKS\,1343$-$601.  The
combined structures in the vicinity of the Norma cluster show a strong
similarity to the Coma cluster in the Great Wall from the CfA redshift
slices by Huchra, Geller and collaborators.  The flow field that
pointed to a so-called Great Attractor does seem explained by what we
now call the Norma Supercluster, together with the galaxy
concentration around PKS\,1343$-$601, as well as the Centaurus Wall,
that stretches from the Pavo cluster ($332\deg,-23\deg$) across the
Galactic Plane to the Centaurus cluster ($320\deg,+22\deg$) at slightly
lower velocities than the Norma Supercluster.

\section*{Acknowledgements}
Discussions with P.A.~Woudt and T.~Jarrett have been invaluable in the
preparation of this review. The contributions of the ZOA team
L. Staveley-Smith (PI), B. Koribalski, P.A. Henning, A.J. Green,
R.D. Ekers, R.F. Haynes, R.M. Price, E.M. Sadler, I. Stewart and
A. Schr\"oder and other participants in the Multibeam HI survey are
gratefully acknowledged. This research used the Lyon-Meudon
Extragalactic Database (LEDA), supplied by the LEDA team at the Centre
de Recherche Astronomique de Lyon, Obs.~de Lyon, and also the
NASA/IPAC Infrared Science Archive (2MASS) and the NASA/IPAC
Extragalactic Database (NED), which are operated by the Jet Propulsion
Laboratory, California Institute of Technology, under contract with
the National Aeronautics and Space Administration. RCKK thanks CONACyT
for their support (research grant 40094F) and the Australian Telescope
National Facility (CSIRO) for their support and hospitality during her
sabbatical.

\subsection*{References}

{\small

\bref
Abell, G.O., Corwin, H.G.Jr., \& Olowin, R.P. 1989,
\apjs 70, 1 

\bref
Balkowski, C., \& Kraan-Korteweg, R.C. (eds.) 1994, 
``Unveiling Large-Scale Structures behind the Milky Way'',
\aspcs 67, (San Francisco: ASP)

\bref
B\"ohringer, H., Neumann, D.M., Schindler, S., \& Kraan-Korteweg R.C. 1996, 
\apj 467, 168

\bref
Cameron, L.M. 1990,
\aaa 233, 16
  
\bref
Donley, J.L., Staveley-Smith, L., Kraan-Korteweg, R.C., et al. 2005,
\aj 129, 220

\bref
Dressler, A., Faber, S.M., Burstein, D., et al. 1987, 
\apj  313, 37 

\bref
Ebeling, H., Mullis, C.R., \& Tully, B.R. 2002,
\apj 580, 774

\bref
Fairall, A.P., \& Kraan-Korteweg, R.C. 2000,
in ``Mapping the Hidden Universe: The Universe behind the Milky Way -
The Universe in HI'', \aspcs 218, 
eds. R.C. Kraan-Korteweg, P.A. Henning \& H. Andernach, 
(San Francisco: ASP), 35

\bref
Fairall, A.P., \& Kraan-Korteweg, R.C. 2005,
in prep. 

\bref
Fairall, A.P., \& Woudt, P.A. (eds.) 2005,
in ``Nearby Large-Scale Structures and the Zone of Avoidance'',
\aspcs 329 (San Francisco: ASP)

\bref
Fairall, A.P., Woudt, P.A., \& Kraan-Korteweg, R.C. 1998, 
\aas 127, 463

\bref 
Henning, P.A., Staveley-Smith, L., Ekers, R.D., et al. 2000,
\aj 119, 2686 

\bref
Hudson, M.J., \& Lynden-Bell, D. 1991, 
\mnras 252, 219

\bref
Hudson, M.J., Smith, R.J., Lucey, J.R., \& Branchini, E. 2004,
\mnras 352, 61

\bref
De Lapparent, V., Geller, M.J., \& Huchra, J.P. 1986,
\apj 302, L1

\bref
Jarrett, T. 2004, 
\pasa 21, 396

\bref
Jarrett, T.-H., Chester, T., Cutri, R., Schneider, S., Rosenberg, J.,
Huchra, J.P., \& Mader, J. 2000a,
\aj 120, 298

\bref
Jarrett, T.-H., Chester, T., Cutri, R., Schneider, S., Skrutskie, M.,
\& Huchra, J.P. 2000b,
\aj 119, 2498 

\bref
Joint IRAS Science Working Group 1988, 
{\sl IRAS} Point Source Catalog,
Version 2 (Washington: US Govt. Printing Office)

\bref
Kerr, F.J., \& Westerhout, G. 1965,
Galactic Structure (Chicago: University of Chicago), 186 

\bref
Kocevski, D.D., Mullis, C.R., \& Ebeling, H. 2004,
\apj 608, 721	

\bref
Kogut, A., Lineweaver, C., Smoot, G.F., et al. 1993,
\apj 419, 1 

\bref
Kolatt, T., Dekel, A., \& Lahav, O. 1995, 
\mnras 275, 797

\bref
Koribalski, B.S., Staveley-Smith, L., Kilborn, V.A., et al. 2004,
\aj 128, 16

\bref
Kraan-Korteweg, R.C. 2000,
\aas 141, 123 
 
\bref
Kraan-Korteweg, R.C., \& Jarrett, T. 2005,
in ``Nearby large-Scale Structures and the Zone of Avoidance'',
\aspcs 329, eds. {A.P. Fairall \& P.A. Woudt}, 
(San Francisco: ASP), 119
(astro-ph/0409391)

\bref
Kraan-Korteweg, R.C., \& Lahav, O. 2000,
\aarv 10, 211

\bref 
Kraan-Korteweg, R.C., \& Woudt, P.A. 1999, 
\pasa 16, 53  

\bref
Kraan-Korteweg, R.C., Loan, A.J., Burton, W.B., Lahav, O.,
Ferguson, H.C., Henning, P.A., \& Lynden-Bell, D. 1994,
\nat 372, 77

\bref 
Kraan-Korteweg, R.C., Fairall, A.P., \& Balkowski, C. 1995, 
\aaa 297, 617

\bref
Kraan-Korteweg, R.C., Woudt, P.A., Cayatte, V., Fairall, A.P., 
Balkowski, C., \& Henning, P.A. 1996, 
\nat 379, 519

\bref 
Kraan-Korteweg, R.C., Schr\"oder, A., Mamon, G., \& Ruphy S. 1998, 
in ``The Impact of Near-Infrared Surveys on 
Galactic and Extragalactic Astronomy'', 
ed. N. Epchtein (Kluwer: Dordrecht), 205

\bref
Kraan-Korteweg R.C., Henning P.A., \& Andernach H. (eds.) 2000,
in ``Mapping the Hidden Universe: The Universe Behind the Milky Way -- 
The Universe in HI'', 
\aspcs 218 (San Francisco: ASP)

\bref
Kraan-Korteweg R.C., Henning P.A., \& Schr\"oder, A.C. 2002,
\aaa 391, 887

\bref
Kraan-Korteweg, R.C., Ochoa, M., Woudt, P.A., \& Andernach, H. 2005a,
in ``Nearby Large-Scale Structures and the Zone of Avoidance'',
\aspcs 329, eds. {A.P. Fairall \& P.A. Woudt}, 
(San Francisco: ASP), 159 (astro-ph/0406044)

\bref
Kraan-Korteweg, R.C., Staveley-Smith, L., Donley, J., Koribalski, B., 
\& Henning, P.A. 2005b,
in ``Maps of the Cosmos'', IAU Symp. 216,
eds. M. Colless \& L. Staveley-Smith, (San Francisco: ASP), in press 
(astro-ph/0311129)

\bref
Lauberts, A. 1982,
The ESO/Uppsala Survey of the ESO (B), (Atlas: ESO, Garching)

\bref
Lucey, J.R., Radburn-Smith, D.J., \& Hudson, M.J. 2005,
in ``Nearby Large-Scale Structures and the Zone of Avoidance'',
\aspcs 329, eds. {A.P. Fairall \& P.A. Woudt}, (San Francisco: ASP),
21 (astro-ph/0412329)

\bref
Lundmark, K. 1940, Lundmark Observatory

\bref
Lynden-Bell, D., \& Lahav, O. 1988,
in ``Large-scale motions in the universe'', 
eds. V.C. Rubin \& G.V. Coyne (Princeton, NJ: Princeton 
University Press), 199

\bref 
McHardy, I.M., Lawrence, A., Pye, J.P., et al. 1981, 
\mnras 197, 893

\bref
Mamon, G.A. 1998,
in ``Wide Field Surveys in Cosmology'', eds. Y. Mellier \& S. Colombi,
(Gif-sur-Yvette: Editions Fronti\`eres), 323

\bref
Meyer, M.J.,  Zwaan, M.A., Webster, R.L., et al. 2004,
\mnras 350, 1195

\bref
Mullis, C.R., Ebeling, H., Kocevski, D.D., \& Tully, R.B. 2005,
in ``Nearby Large-Scale Structures and the Zone of Avoidance'',
\aspcs 329, eds. {A.P. Fairall \& P.A. Woudt} (San Francisco: ASP),
183 

\bref
Nagayama, T., Ph.D. thesis, Nagoya University, 2004

\bref
Nagayama, T., Woudt, P.A., Nagashima, C., et al. 2004,
\mnras 354, 980

\bref
Nagayama, T., Nagata, T., Sato, S., Woudt, P.A., \& IRSF/SIRIUS team 2005,
in ``Nearby Large-Scale Structures and the Zone of Avoidance'',
\aspcs 329, eds. {A.P. Fairall \& P.A. Woudt} (San Francisco: ASP),
177

\bref
Nilson, P. 1973, 
Uppsala General Catalog of Galaxies,
(Uppsala: University of Uppsala)

\bref
Paturel, G., Petit, C., Rousseau, J., \& Vauglin, I. 2003,
\aaa 405, 1

\bref
Peebles, P.J.E. 1994,
\apj 429, 43 

\bref
Rousseau, J., Paturel, G., Vauglin, I., Schr\"oder, A., et al. 2000,
\aaa 363, 62

\bref
Salem, C., \& Kraan-Korteweg, R.C., in prep.

\bref
Saunders, W., D'Mellow, K.J., Valentine H., et al. 2000,
in ``Mapping the Hidden Universe: The Universe behind the Milky Way -
The Universe in HI'', 
\aspcs 218, eds. R.C. Kraan-Korteweg, P.A. Henning \&
H. Andernach, (San Francisco: ASP), 141

\bref
Schlegel, D.J., Finkbeiner, D.P., \& Davis M. 1998,
\apj 500, 525

\bref
Schr\"oder, A., Kraan-Korteweg, R.C., Mamon, G.A., \& Ruphy S. 1997,
in ``Extragalactic Astronomy in the Infrared'', 
eds. G.A. Mamon, T.X. Thuan \& J. Tran Thanh Van 
(Editions Fronti\`eres: Gif-sur-Yvette), 381

\bref 
Schr\"oder, A., Kraan-Korteweg, R.C., \& Mamon G.A. 1999,
\pasa 16, 42

\bref
Schr\"oder, A., Kraan-Korteweg, R.C., \& Mamon, G.A. 2000,
in ``Mapping the Hidden Universe: The Universe behind the Milky Way -
The Universe in HI'', 
\aspcs 218, eds. R.C. Kraan-Korteweg, P.A. Henning \&
H. Andernach, (San Francisco: ASP), 119

\bref
Schr\"oder, A., Kraan-Korteweg, R.C., Mamon, G.A., \& Woudt, P.A. 2005,
in ``Nearby Large-Scale Structures and the Zone of Avoidance'',
\aspcs 329, eds. {A.P. Fairall \& P.A. Woudt} (San Francisco: ASP),
167
(astro-ph/0407019)

\bref
Schr\"oder, A.C., Kraan-Korteweg, R.C., \& Henning, P.A., in prep.

\bref
Staveley-Smith, L., Wilson, W.E., Bird, T.S., et al. 1996,
\pasa 13, 243

\bref
Tonry, J.L., \& Davis, M. 1981,
\apj 246, 680

\bref
Vauglin, I., Rousseau, J., Paturel, G., et al. 2002,
\aaa 387, 1

\bref Vorontsov-Velyaminov, B., Archipova, V.P., \& Krasnogorskaja,
A. 1962-1974, 
Morphological Catalogue of Galaxies, Vol.~I-V (Moscow: Moscow State 
University)

\bref
Wakamatsu K., Parker Q.E., Malkan M., \& Karoji H. 2000,
in ``Mapping the Hidden Universe: The Universe behind the Milky Way -
The Universe in HI'', \aspcs 218, 
eds. R.C. Kraan-Korteweg, P.A. Henning \& H. Andernach, 
(San Francisco: ASP), 187

\bref
Wakamatsu, K., Malkan, M.A., Nishida, M.T., Parker, Q.A., Saunders, W., 
\& Watson, F.G. 2005,
in ``Nearby Large-Scale Structures and the Zone of Avoidance'', 
\aspcs 329, 
eds. A.P. Fairall \& P.A. Woudt, (San Francisco: ASP), 149
  
\bref
Woudt, P.A. 1998, 
Ph.D. thesis, Univ. of Cape Town

\bref
Woudt, P.A., \& Kraan-Korteweg, R.C. 2001,
\aaa 380, 441

\bref
Woudt P.A., Kraan-Korteweg, R.C., \& Fairall, A.P. 1999, 
\aaa 352, 39

\bref
Woudt, P.A., Kraan-Korteweg, R.C., \& Fairall, A.P. 2000,
in ``Mapping the Hidden Universe: The Universe behind the Milky Way -
The Universe in HI'', \aspcs 218,
eds. R.C. Kraan-Korteweg, P.A. Henning \& H. Andernach, 203

\bref
Woudt, P.A., Kraan-Korteweg, R.C., Cayatte, V., Balkowski, C., 
\& Felenbok, P. 2004,
\aaa 415, 9

\vfill

\end{document}

%% file: abb.tex
\def\la{\mathrel{\hbox{\rlap{\hbox{\lower4pt\hbox{$\sim$}}}\hbox{$<$}}}}
\def\ga{\mathrel{\hbox{\rlap{\hbox{\lower4pt\hbox{$\sim$}}}\hbox{$>$}}}}

\def\arcmin{\hbox{$^\prime$}}
\def\arcsec{\hbox{$^{\prime\prime}$}}

\def\fm{\hbox{$.\!\!^{\rm m}$}}

\def\fdg{\hbox{$.\!\!^\circ$}}
\def\farcm{\hbox{$.\mkern-4mu^\prime$}}

\def\micron{\hbox{$\mu$m}}

\newcommand{\etal}{{et~al.}\,}      
\def\deg{{^\circ}}

\newcommand{\kms}{{\,km\,s$^{-1}$}}

\newcommand{\B}{{$B$}}

\newcommand{\II}{{$I_c$}}
\newcommand{\I}{{$I$}}
\newcommand{\J}{{$J$}}
\newcommand{\HH}{{$H$}}
\newcommand{\K}{{$K_s$}}

\def\apj    {{ApJ }}

\def\apjs   {{ApJS }}
\def\aj     {{AJ }}
\def\aaa    {{A\&A }}

\def\aas    {{A\&ASS }}

\def\aarv   {{A\&ARv }}

\def\mnras  {{MNRAS }}
\def\nat    {{Nature }}

\def\pasa   {{PASA }}

\def\aspcs  {{ASP Conf. Ser. }}